\documentclass[preprintnumbers, superscriptaddress, reprint, longbibliography, aps, prb, floatfix]{revtex4-2}
\usepackage{amsmath}
\usepackage{amssymb}
\usepackage{mathrsfs}
\usepackage{graphicx}
\usepackage{dcolumn}
\usepackage{bm}
\usepackage{braket}
\usepackage{xcolor}

\usepackage[colorlinks=true, linkcolor=blue, citecolor=blue, urlcolor=blue]{hyperref}

\begin{document}
\title{Edge-driven transition between extended quantum anomalous Hall crystal and fractional Chern insulator in rhombohedral graphene multilayers}

\author{Zezhu Wei}
\affiliation{Theoretical Division, T-4, Los Alamos National Laboratory, Los Alamos, New Mexico 87545, USA}

\author{Ang-Kun Wu}
\affiliation{Theoretical Division, T-4, Los Alamos National Laboratory, Los Alamos, New Mexico 87545, USA}

\author{Miguel Gonçalves}
\affiliation{Theoretical Division, T-4, Los Alamos National Laboratory, Los Alamos, New Mexico 87545, USA}

\author{Shi-Zeng Lin}
\affiliation{Theoretical Division, T-4 and CNLS, Los Alamos National Laboratory, Los Alamos, New Mexico 87545, USA}
\affiliation{Center for Integrated Nanotechnologies (CINT), Los Alamos National Laboratory, Los Alamos, New Mexico 87545, USA}

\begin{abstract}
Fractional Chern insulators (FCI) with fractionally quantized Hall conductance at fractional fillings and an extended quantum anomalous Hall (EQAH) crystal with an integer quantized Hall conductance over an extended region of doping were recently observed in pentalayer graphene. One particularly puzzling observation is the transition between the EQAH and FCI regimes, driven either by temperature or electrical current. Here we propose a scenario to understand these transitions based on the topologically protected gapless edge modes that are present in both the FCI and EQAH phases and should be most relevant at temperature scales below the energy gap. Our consideration is based on the simple assumption that the edge velocity in FCI is smaller than that in EQAHE and thus contributes to a higher entropy. We further argue that domains with opposite fractionally quantized Hall conductance are ubiquitous in the devices due to disorder, which gives rise to a network of edge modes. The velocity of the edge modes between domains is further reduced due to edge reconstruction. The edge velocity can also be reduced by current when the occupation of the edge mode approaches the gap edge. The edge entropy therefore drives the transition from EQAH to FCI either by temperature or current at a nonzero temperature. 
\end{abstract}
\date{\today}
\maketitle

\section{Introduction}
FCI is a topological state that emerged in strongly correlated topological flat bands. When the topological flat band mimics a Landau level, i.e., the Berry curvature is uniform, and the trace condition is satisfied, partial filling of the topological flat band stabilizes the FCI \cite{PhysRevLett.106.236802,PhysRevLett.106.236803,PhysRevLett.106.236804,sheng2011fractional,PhysRevX.1.021014,PhysRevB.90.165139,PhysRevResearch.2.023237,PhysRevLett.127.246403,PhysRevResearch.5.023167,PhysRevB.108.205144}. FCI has the same topological order as the fractional quantum Hall effect (FQHE) in a two-dimensional electron gas under a strong magnetic field, but the difference is that the former occurs in topological bands without the need for an external magnetic field. The FCI was proposed more than a decade ago in several toy models \cite{PhysRevLett.106.236802,PhysRevLett.106.236803,PhysRevLett.106.236804,sheng2011fractional,PhysRevX.1.021014}. The advent of moir\'e superlattice eventually led to the experimental breakthrough of the observation of FCI in twisted MoTe$_2$ homobilayer \cite{Park2023,PhysRevX.13.031037,Cai2023,Zeng2023}, consistent with the theoretical calculations \cite{PhysRevResearch.3.L032070,PhysRevB.108.085117,PhysRevLett.132.096602}. This discovery was soon followed by the experimental observation of FCI in pentalayer graphene \cite{Lu2024} and other rhombohedral graphene multilayers \cite{Lu_Han_Yao_Yang_Seo_Shi_Ye_Watanabe_Taniguchi_Ju_2024,xie2024evenodddenominatorfractionalquantum,Waters_Okounkova2024}.

Theoretically, the FCI in MoTe$_2$ can be understood in terms of the partially filling of a topological flat band that resembles a Landau level \cite{PhysRevLett.122.086402,PhysRevLett.132.096602,li2024contrasting}. The FCI in pentalayer graphene, however, poses a significant challenge for the theoretical modeling of this system. The noninteracting band structure with an experimental setup with a weak moir\'e potential is a gapless metal. The Hartree--Fock calculations show a gap opening at integer filling and stabilize a Chern insulator with an integer quantized Hall conductance. The exact diagonalization calculations find FCI by partial filling of the Hartree-Fock band obtained at integer filling \cite{zhou2023fractionalquantumanomaloushall,dong2023anomaloushallcrystalsrhombohedral,tan2024parentberrycurvatureideal,Dong_Patri_Senthil_2024,soejima2024anomaloushallcrystalsrhombohedral,PhysRevB.109.205122,Kwan_Yu_Herzog-Arbeitman_Efetov_Regnault_Bernevig_2023,Yu_Herzog-Arbeitman_Kwan_Regnault_Bernevig_2024,PhysRevB.110.075109}. However, the justification for this procedure is unclear.

Despite the challenge in modeling, recent experiments have observed even more intriguing behavior. New experimental progress in cooling now allows one to measure the devices at temperatures down to 10 mK. Surprisingly, upon cooling to a lower temperature, the FCI observed initially at fractional filling at about 100 mK in Ref.~\cite{Lu2024} was replaced by a state with integer quantized Hall conductance, which was called extended quantum anomalous Hall crystal in Ref.~\cite{Lu_Han_Yao_Yang_Seo_Shi_Ye_Watanabe_Taniguchi_Ju_2024}. This EQAH regime is stable over a wide range of doping from $\nu=0.5$ to $\nu=1$. Furthermore, the EQAH regime is stable upon applying a small current up to a threshold current of the order of about $0.5$ nA, beyond which the system switches to an FCI state at the corresponding filling fraction. The differential conductance resembles that for an $s$-wave superconductor. The experimental observation of current- and temperature-driven change from EQAH to FCI has motivated theoretical efforts to understand this phenomenology \cite{Sarma_Xie_2024,Patri_Dong_Senthil_2024} and is the main purpose of the current work.

The nature of EQAH is not clear now, and several possibilities have been suggested.  One possibility is that electrons crystallize in a crystal similar to a Wigner crystal, but now with quantized Hall conductance \cite{Patri_Dong_Senthil_2024}. This new form of electron crystal is called quantum anomalous Hall crystal in literature \cite{zhou2023fractionalquantumanomaloushall,dong2023anomaloushallcrystalsrhombohedral,tan2024parentberrycurvatureideal,sheng2024quantumanomaloushallcrystal,Dong_Patri_Senthil_2024,PhysRevLett.132.236601,soejima2024anomaloushallcrystalsrhombohedral}, and this picture has been supported by several theoretical calculations based on low energy model for pentalayer graphene. The second possibility is to start from the Chern insulator at $\nu=1$ and consider fractional filling at $\nu<1$ as the hole doping of the Chern insulator. If the doped holes form a topological trivial Wigner crystal or are Anderson localized due to impurities, then the resulting state still has integer Hall conductance in the presence of hole doping. We will not attempt to address the nature of the EQAH in this work. One implication of the observed transition from EQAH to FCI is that for the range of displacement field where it takes place, EQAH and FCI have very close energy, and the ground state energy density for EQAH is slightly smaller than that of FCI.

As a microscopic many-body treatment of EQAH to FCI transition is not feasible using an appropriate Hamiltonian for pentalayer graphene at this moment, we propose a plausible scenario to account for the EQAH to FCI transition. Our picture is based on the topologically protected edge modes in EQAH and FCI. When temperature $T$ is much smaller than the gap, since the bulk is fully gapped, the gapless edge mode becomes important for the observed transition of EQAH to FCI. %
On the other hand, the edge modes of the FCI have much smaller edge velocity than that of the EQAH state. As edge modes with smaller velocities provide a higher entropy contribution, the corresponding state is favored at a higher temperature \cite{PhysRevB.85.165134,giamarchi2003quantum}.
The application of current can also enhance the entropy as the filling of the edge mode approaches the edge of the gap, where the edge mode velocity necessarily becomes smaller, see Fig.~\ref{f1}. This entropy enhancement can also happen when the edge mode is filled to local extrema.

In the thermodynamic limit, for a uniform domain, the edge contribution is subleading compared to the bulk. In the experiment, the typical device size is about a few micrometers, and the average electron distance is about 10 nm. So, typically the device contains of the order of $100\times 100$ unit cells. For such a small system, the edge contribution may be important, particularly given that the ground state energies for EQAH and FCI are close. Another important ingredient is disorder, which is inevitably present in devices. In 2D, disorder causes the formation of domains across devices whose size is controlled by competition between the elastic and disorder energies. These domains with opposite Hall conductance are present and are responsible for the experimentally observed hysteresis in Hall conductance when the magnetic field sweeps. The formation of domains in the 2D moir\'e is ubiquitous and has been imaged directly in twisted bilayer graphene \cite{Grover2022Mosaic}, where a mosaic of Chern domains has been observed. The domains between different FCI states in twisted MoTe$_2$ were imaged in Ref.~\cite{ji2024localprobebulkedge}. More relevant to our discussion is the tetralayer graphene, where superconductivity has been observed \cite{Han_Lu_Yao_Shi2024}. The fluctuations of resistance as a function of time in the superconducting state were ascribed to the fluctuations of superconducting domains with opposite orbital magnetization and valley index. In the domain walls, we will show that the edge velocity is further reduced due to the edge reconstruction, which further enhances the entropy contribution.

The paper is organized as follows. In Sec.~\ref{sec2}, we discuss the many-body gap and the edge velocity in EQAH and FCI. In Sec.~\ref{sec3}, we discuss the entropy contribution due to the gapless edge mode. In Sec.~\ref{sec4}, we propose a mechanism for the temperature- and current-driven transition from EQAH to FCI based on the edge contribution. In Sec.~\ref{sec5}, we discuss domain formation and edge velocity reduction in devices with disorders. The paper is concluded by a brief discussion and summary in Sec.~\ref{sec6}. Throughout the paper, we will take $\hbar=k_B=c=e=1$ if we do not write them explicitly.

\section{gap size and edge velocity in EQAH and FCI}\label{sec2}
We first discuss the many-body gap of EQAH and FCI by making analogies to the FQHE.  For the non-interacting Landau level, the gap is given by the cyclotron frequency $\hbar \omega_c$ with $\omega_c=eB/m c$, where $B$ is the external magnetic field and $m$ is the electron mass. The FQAHE can be regarded as Landau levels of weakly interacting composite fermions. For the Jain sequence at filling $\nu=\rho\phi_0/B=n/(2pn+1)$ \cite{PhysRevLett.63.199,jainthirty}, the effective magnetic field for the composite fermion is $B^*=B-2p\rho\phi_0$ with integers $p$ and $n$, electron density $\rho$, and {flux quantum} $\phi_0=hc/e$. Therefore, the gap of FQAHE is smaller for a larger Jain index $n$ and becomes gapless at half-filling corresponding to $n=\infty$ and $p=1$. The gap size versus $n$ has been verified experimentally, which shows great agreement \cite{PhysRevLett.127.056801,PhysRevB.106.L041301}. In FCI, the topological flat band deviates from the Landau level. As a consequence, the residual interaction between the composite fermions remains strong \cite{PhysRevB.86.165129,PhysRevLett.132.236502}. It is natural that the scaling of the gap size with $n$ breaks down. Indeed, the gaps for FCI in pentalayer graphene extracted from transport measurement are similar for all fractional filling \cite{Lu2024,PhysRevB.109.L241115}. We remark that similar many-body gap size versus filling in the FCI is obtained in a toy model study of FCI, as discussed in detail in Appendix~\ref{appA}. Nevertheless, the key to the following discussion is that the gap size of the FCI is smaller than that of the Chern insulator, as observed in the experiment.

To the best of our knowledge, the edge mode velocity in FCI has not been calculated or measured. A crude estimate of the edge velocity is given by $\Delta/G$, where $\Delta$ is the size of the gap and $G$ is the reciprocal wave vector. This estimate shows that the edge velocity $v_e$ for the higher Jain state $n$ in FQHE is smaller. This is consistent with the numerical calculation of edge velocity in FQHE in Ref.~\cite{PhysRevB.80.235330}. This estimate also implies that the edge velocity in FCI is smaller than that in EQAH, $v_{\mathrm{FCI}}<v_{\mathrm{EQAH}}$. Of course, $v_e$ is not universal and also depends on the details of the edge, i.e., the shape of the devices. For simplicity, we neglect the non-universal details and assume $v_{\mathrm{FCI}}<v_{\mathrm{EQAH}}$ in the following discussion.

\section{Entropy of the edge mode} \label{sec3}
When $T<\Delta$, the entropy mainly originates from the gapless edge modes because the bulk is fully gapped. Phonon of the EQAH crystal is also gapped due to the pinning by disorders. Here we discuss the contribution of entropy from a chiral edge mode with velocity $v_e$ both in EQAH and FCI. 

For the EQAH, one can use the Hartree--Fock band structure. For a stripe geometry, the chiral edge mode along the edge with translation invariance is given by $\mathcal{H}=v_e(k_F) (k-k_F) c_k^\dagger c_k$. Here we allow the edge velocity to be $k$ dependent. When the edge dispersion $\epsilon_e(k)$ has extrema, the corresponding $v_e(k)$ vanishes. $v_e(k)$ is also significantly reduced at the chemical potential when the edge mode merges into the bulk state as a result of level repulsion between bulk and edge modes; see Fig.~\ref{f1}(b) for an example. At a low temperature $T$, when $\epsilon_e(k)\approx v_e(k_F) (k-k_F)$---accurate over the temperature window---the entropy of the edge mode with length $L$ is 
\begin{align}\label{eq1}
    S_e= \frac{\pi T L}{6 v_e}.
\end{align}

The edge physics in FCI can be deduced based on parton theory. Here we take $\nu=1/3$ as an example \cite{PhysRevB.85.165134}. The electron operator can be written as $c(r)=f_1(r)f_2(r)f_3(r)$, where the parton $f_i$ carries the U(1) charge $e/3$. The parton construction necessarily introduces emergent gauge fields. Standard mean-field decoupling can be performed to obtain a mean-field Hamiltonian for the parton, $\mathcal{H}_f=f_i^\dagger M_{ij} f_j$, which produces the band structure for $f_i$. We need to enlarge the unit cell of $\mathcal{H}_f$ to be three times larger than the original unit cell, such that $f_i$ fully fills the band. To describe FCI, we demand $f_i$ to fully fill the lowest $C=1$ bands. Going back to the physical Hilbert space, we need to impose the constraint $c^\dagger c=f_i^\dagger f_i$ for any $i$. The FCI ground state wave function can be obtained by projecting out the unphysical components of the $\mathcal{H}_f$ ground state wavefunction, i.e., $\Phi_{\text{FCI}}=\bra{0}\prod_{j=1}^N f_1(r_j)f_2(r_j)f_3(r_j) \ket{\Phi_{MF}}$, where $\bra{0}$ is the parton vacuum. $f_i$ breaks the translation symmetry of the original lattice, but the symmetry is restored for the physical fermion $c$ after the projection in the physical Hilbert space. Or, in other words, the translation symmetry acts projectively on $f$ due to the emergent gauge redundancy in the parton construction. Then the entropy for the edge mode in FCI can be obtained in terms of $f$ partons, similar to that in EQAH. The gauge fields in EQAH and FCI are gapped, and their contribution to entropy can be safely neglected at a low $T$.  

The edge entropy contribution for EQAH and FCI can also be obtained using bosonization for the chiral Luttinger liquid. The edge Hamiltonian for the bosonic field is 
\begin{equation}
	H_{e} = \sum_{k>0}  v_{e} k b_{k}^{\dagger} b_{k} = \frac{v_{e}}{4\pi} \int dx\, (\partial_{x} \varphi)^2,
\end{equation}
with interaction renormalized edge velocity $v_e$, and edge bosonic field $\varphi$ and the corresponding operator $b_k$. Straightforward calculations also yield the edge entropy in Eq. \eqref{eq1}~\cite{giamarchi2003quantum}.

\section{Temperature- and current-driven transition from EQAH to FCI}\label{sec4}

Now we are in a position to discuss the temperature- and current-driven transition from EQAH to FCI based on the edge entropy. Here we consider a square geometry with linear size $L$. We assume that there is a single domain of EQAH or FCI in the device, valid in a small device. For a large device, multiple domains are inevitably induced by disorder or temperature, which will be discussed in detail in Sec.~\ref{sec5}. The free energy difference between the EQAH and FCI is
\begin{align}
    \Delta F=(\mathcal{E}_{\rm EQAH}-\mathcal{E}_{\rm FCI})L^2- \left(\frac{1}{v_{\rm EQAH}}-\frac{1}{v_{\rm FCI}}\right)\frac{2\pi LT^2}{3},
\end{align}
where $\mathcal{E}_{\rm EQAH}$ and $\mathcal{E}_{\rm FCI}$ are the ground state energy density for EQAH and FCI, respectively. Here we consider FCI with only one edge mode, such as the FCI at $\nu=1/3$. For the FCI with more than one edge mode, {such as the FCI at $\nu=2/3$ with two counter-propagating edge modes, we need to sum over all the edge contributions}. These edge modes velocity can be renormalized due to the interactions between the modes; see Sec.~\ref{sec5} for more detailed discussions. For a small device size, the FCI state has lower free energy when $T_c>\sqrt{{3L(\mathcal{E}_{\text{EQAH}}-\mathcal{E}_{\text{FCI}})}/{2\pi(v_{\text{EQAH}}^{-1}-v_{\text{FCI}}^{-1})}}$. The $T$-driven transition is possible only when $T_c$ is smaller than the FCI and EQAH gaps. Therefore, for a large device with a single domain, this transition may not happen. 

We estimate the edge entropy contribution using the parameters extracted from Ref.~\cite{Lu2024}. We take the gap to be $1$ meV and the Brillouin size to be $G=2\pi/11.5\ \mathrm{nm}^{-1}$. This gives a rough estimate of the edge velocity $v_e\sim 3000\ \mathrm{m/s}$. For a device of size $L=3\ \mathrm{\mu m}$ and at $T=0.1$ K, we estimate the edge entropy contribution to the free energy is $T S_e\sim 0.25$ meV. When $\mathcal{E}_{\text{EQAH}}$ and $\mathcal{E}_{\text{FCI}}$ are close, such an edge entropy contribution can be enough to drive the transition from EQAH and FCI as observed in the experiment.   

\begin{figure}[t]
  \begin{center}
  \includegraphics[width=\columnwidth]{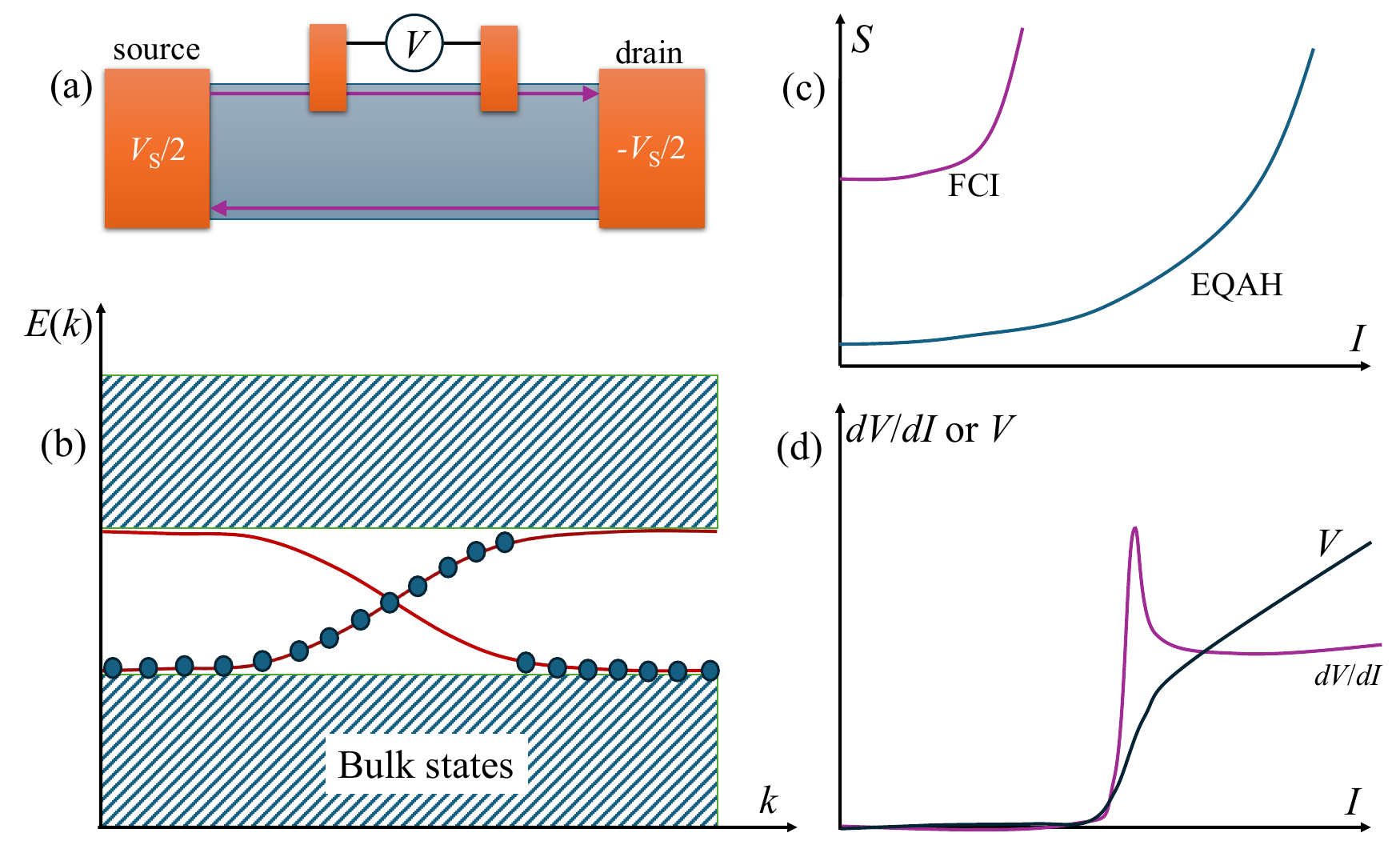}
  \end{center}
\caption{{(a) Electrode configuration for the transport measurement in EQAH and FCI, where current is injected through source and drain biased at voltage $\pm V_S/2$. The longitudinal resistance is probed by the electrodes in the upper edge. (b) Schematic view of the occupation of the edge mode in the presence of a current. (c) Sketch of the edge entropy versus current in the EQAH and FCI. The edge entropy increases significantly when the edge states near the gap edge start to be populated at a threshold current. (d) Sketch of the $I$-$V$ and $dV/dI$ curves when the system switches from EQAH at low $I$ to FCI at high $I$.} %
} 
  \label{f1}
\end{figure}

We then turn to the current-driven transition. To be concrete, we consider a setup where the sample is attached to electrodes with bias voltage $\pm V_S/2$ to inject current into the device, as sketched in Fig.~\ref{f1}(a). Then the bottom (top) edge is equilibrated with the right (left) electrode, such that the electron occupation for the left (right) moving branch is pushed down (up). This results in a current $I=V_S e^2/h $. When particle-hole symmetry is present, the total edge energy remains the same in the presence of the current. The violation of  particle-hole symmetry makes the energy depend on current, which we will not consider here for simplicity. The maximum dissipationless current that the edge can carry is about $I_c\approx \Delta e/h$. Beyond $I_c$, the extra current is carried by the bulk states and becomes dissipative. In the presence of current $I<I_c$, $v_e$ depends on $I$ as the occupation of the edge modes changes. As $I$ approaches $I_c$, $V_S$ gets close to the edge of the gap, and $v_e$ is reduced due to the level repulsion between the edge and bulk modes and also because of the periodicity of the Brillouin zone, see Fig.~\ref{f1}(b). Then the edge entropy increases with current, as sketched in Fig.~\ref{f1}(c). An explicit calculation using the Haldane model is presented in Appendix \ref{appB}, which shows that the edge entropy is significantly enhanced for a voltage near the gap edge. We again need to compare the edge entropy for EQAH and FCI in the presence of current. Since the gap for FCI is smaller than EQAH, when one ramps up $I$, {the electric potential for the FCI edge state first reaches the gap edge, which further enhances the edge entropy as shown in Fig.~\ref{f1}(c).} As a consequence, a transition from EQAH to FCI can be triggered by the current. The threshold current observed in the experiment is about 0.5 nA, which corresponds to an FCI gap of the order of 1.3 K. This value roughly agrees with the gap size extracted from transport measurement ($\Delta \sim 5$ K) \cite{PhysRevB.109.L241115}. For a general configuration of electrodes, the edge energy also depends on the current, but the sharp increases of edge entropy near the gap edge should be dominant over the edge energy change.

{For $I<I_c$, the current is carried mainly by the chiral edge mode of EQAH, so the longitudinal resistance is small at low temperature. The longitudinal resistance can be measured by attaching electrodes at the upper edge, see Fig.~\ref{f1}(a). Once at $I\approx I_c$, the system switches to FCI, with the chiral edge mode almost fully filled. With temperature, the thermally excited bulk starts to contribute, and as a consequence, there is a jump (but rounded by temperature) in the longitudinal resistance. As $I$ increases further, the transport is dominated by the bulk and the longitudinal resistance saturates to a bulk value. The longitudinal $I$-$V$ and $dV/dI$ curves in this switch process are sketched in Fig.~\ref{f1}(d), which are consistent with the experiments \cite{Lu_Han_Yao_Yang_Seo_Shi_Ye_Watanabe_Taniguchi_Ju_2024}.}

\section{Domain formation due to disorder}\label{sec5}

In this section, we discuss the formation of domains due to disorder and the associated edge mode entropy for both the EQAH and FCI states. 
A phase diagram consisting of several domain-dominated phases will be given at the end of this section.

\begin{figure}[htbp]
	\centering
	\includegraphics[width=\columnwidth]{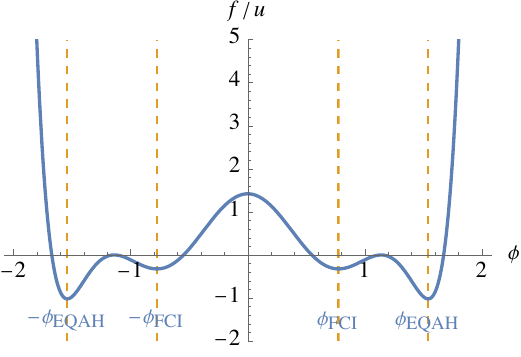}
	\caption{Free energy density profile for a uniform order parameter $\phi$ when $s=0.5u$, $r=3u$, and $\Phi_{0}=1.3$. $-\phi_{\text{FCI}}$ and $\phi_{\text{FCI}}$ are FCI states with an opposite Hall conductance, and $-\phi_{\text{EQAH}}$ and $\phi_{\text{EQAH}}$ are EQAH states with an opposite Hall conductance.}
 \label{fig:free_energy}
\end{figure}

We describe the system phenomenologically through a free energy density $f$.
At zero temperature, $f$ is given by
\begin{equation}
	f(\phi) = - \frac{r}{2} (\phi^2 - \Phi_0)^2 + \frac{c}{2} (\nabla \phi)^2 - s (\phi^2 - \Phi_0)^3 
	+ u (\phi^2 -\Phi_0)^4,
\end{equation}
where $r,s,u,c,\Phi_{0}>0$ and $\phi$ is the order parameter, {corresponding to orbital magnetization in the experiment.} 
As shown in Fig.~\ref{fig:free_energy}, uniform phases in which $\phi$ does not vary with $x$ are given by the following order parameters that minimize $f$:
\begin{align}
	\pm \phi_{\text{FCI}} & = \pm \sqrt{\varphi_{-}+\Phi_0}, \\
	\pm \phi_{\text{EQAH}} & = \pm \sqrt{\varphi_{+}+\Phi_0},
\end{align}
where $\varphi_{\pm} = \frac{3s}{8u} \pm \frac{1}{2} \sqrt{\frac{r}{u}+ \left(\frac{3s}{4u}\right)^2}$ are locations of the two minimums of the auxiliary function $\mathcal{F}(\varphi) = - \frac{r}{2} \varphi^2 - s \varphi^3 + u \varphi^4$.
$\pm \phi_{\text{FCI}}$  represent the FCI phases, and $\pm \phi_{\text{EQAH}}$  represent the EQAH phases.
To have physical $\pm \phi_{\text{FCI}}$ phases, we require $\varphi_{-}+\Phi_{0}>0$. 
The free energy density of $\pm \phi_{\text{FCI}}$ is $\mathcal{F}(\varphi_{-})$ and is greater than $\mathcal{F}(\varphi_{+})$, which is the free energy density of $\pm \phi_{\text{EQAH}}$. 

If the stiffness term $c (\nabla \phi)^2 / 2$ in the free energy density $f(\phi)$ is dominating or the potential wells for the stable phases are shallow, the order parameter of two neighboring domains tends to change continuously from one domain to another across the domain wall, over a distance of the linear domain size $L$. 
We will focus on this smooth domain wall regime in the main text. 
The opposite regime, in which the non-linear terms dominate over the stiffness term and domain walls become sharp, will be discussed in Appendix~\ref{appC}.

Now we address the smooth domain wall regime. If the order parameters of the neighboring domains are $\phi_{a}$ and $\phi_{b}$, then the cost of the free energy density by forming a domain wall can be estimated as
\begin{equation}
	f_{\text{DW}} = \frac{c}{2} \left(\frac{2\Delta\phi}{L}\right)^2,
\end{equation}
where $\Delta\phi=(\phi_{b}-\phi_{a})/2$, and we have ignored the contribution from nonlinear terms in the free energy.

The EQAH and FCI phases with opposite Hall conductance have opposite orbital magnetization. We model the disorder in the devices as random magnetic fields $b(\mathbf{x})$ coupled to the orbital magnetization. This coupling can also be written in terms of the order parameter field directly
\begin{align}
	F_{\text{dis}} = \int d^{2}\mathbf{x} \, b(\mathbf{x}) \phi(\mathbf{x}),
\end{align}
and we consider a short-range random field with zero mean and variance given by
\begin{equation}
	\overline{b(\mathbf{x}) b(\mathbf{x}')} = h^2 \delta^{(2)}( \mathbf{x}-\mathbf{x}' ),
\end{equation}
where $h$ is the disorder strength.
The system can have free energy gain from the disorder by optimizing the domain size, and we will estimate this energy gain following the approach used in Ref.~\cite{thomson2021:PhysRevB.103.125138}.
Within a domain of linear length $L$ and order parameter $\phi$, its mean energy is zero, but the root mean square energy is $F_{\text{rms}} = \phi h L$. 
The energy of a single domain fluctuates around zero within a typical range of $\pm F_{\text{rms}}$.
For a system containing domains between $\phi_{a}$ and $\phi_{b}$, assuming $\phi_{a}<\phi_{b}$, it can reduce its total energy 
by choosing a typical energy $+F_{\text{rms}}$ for domains of $\phi_{a}$ and a typical energy $-F_{\text{rms}}$ for domains of $\phi_{b}$. 
The average energy density gain per domain due to disorder is estimated to be
\begin{equation}
	f_{\text{dis}} \approx \frac{1}{L^2} \frac{1}{2} (\phi_{a}-\phi_{b}) h  L
	= - (\Delta \phi) h  \frac{1}{L}.
\end{equation}

At a finite temperature, domains can further gain free energy through the entropy of topological edge modes in the domain walls~\cite{shavit2022:PhysRevLett.128.156801}.
For a domain wall of length $\ell$ with $N$ co-propagating edge channels, the entropy is given by $S =(\pi T \ell/6) \sum_{i=1}^{N}{v_i^{-1}}$. We focus on the case of $N=2$ for concreteness. 
Interactions between multiple edge modes would renormalize the edge velocity~\cite{WenBook,ferraro2017:10.21468/SciPostPhys.3.2.014,wei2024:PhysRevB.110.075306}, resulting in a larger edge mode entropy.
For two co-propagating edge modes with a density-density interaction between them, we can write down
the following Lagrangian density:
\begin{equation}
    \begin{split}
        \mathcal{L} = {} &  \frac{  \partial_{x} \varphi_{a} (\partial_{t}-v_{a} \partial_{x}) \varphi_{a}}{4\pi}  + \frac{  \partial_{x} \varphi_{b} (\partial_{t}-v_{b} \partial_{x}) \varphi_{b}}{4\pi}  
	\\
 &- \frac{ w}{2\pi} (\partial_{x} \varphi_{a})(\partial_{x}\varphi_{b}),
    \end{split}
\end{equation}
where $v_{a,b}$ are the bare velocities, $w$ is the interaction strength, and $\varphi_{a,b}$ are the chiral bosonic fields at the edge. The velocities of the eigenmodes are renormalized to
\begin{align}\label{eq:eigen_velocity}
	v_{\alpha} &= \frac{1}{2} \left[ v_{a} + v_{b} + \sqrt{(v_{a}-v_{b})^2+4 w^2}  \right],\\
	v_{\beta} &= \frac{1}{2} \left[v_{a} + v_{b} - \sqrt{(v_{a}-v_{b})^2+4 w^2}\right]  .
\end{align}
When $v_{a}=v_{b}=v$, the eigenmode velocities are $v_{\alpha} = v + w$ and $v_{\beta} = v -w  $.
Then the entropy of the domain wall is increased by a factor of $v^2/(v^2-w^2)$,
\begin{align}
	S = \frac{\pi T \ell}{6} \left(\frac{1}{v+w} + \frac{1}{v-w}\right)
	= \frac{\pi T \ell}{3v} \frac{v^2}{v^2-w^2} .
\end{align}
We can find the free energy density contribution as
\begin{equation}
	f_{\text{ent}} = - \frac{TS}{L^2} = - \frac{T}{L^2}   \frac{\pi T }{3v} (2L) \frac{v^2}{v^2-w^2} =  -\frac{2 \pi T^2 }{3 v L}  \frac{v^2}{v^2 - w^2}.
\end{equation}
In passing, we would like to remark that for counter-propagating edge modes with velocities $v_{a,b}$ and interaction strength $w$, the renormalized velocities are
\begin{align}
	v_{\alpha} &= \frac{1}{2} \left[ v_{a} - v_{b} + \sqrt{(v_{a}+v_{b})^2-4 w^2}  \right],\\
	v_{\beta} &= \frac{1}{2} \left[-v_{a} + v_{b} + \sqrt{(v_{a}+v_{b})^2-4 w^2}\right]  .
\end{align}
This case would be relevant for FCI or domain walls with counter-propagating edge modes.
The edge mode entropy also increases after velocity renormalization.

For completeness, we should also consider the entropy contribution from rearranging the domains.
When there are $\mathcal{N}$ domains, the number of possible arrangements is $2^{\mathcal{N}}$, and hence the entropy per domain is $\log(2^{\mathcal{N}})/\mathcal{N}=\log{2}$. Due to the presence of the
disorder field $b$, which can pin the domains, the entropy is expected to be even lower than $\log{2}$ and eventually becomes negligible.

To simplify the analysis, we temporarily neglect the interaction $w$ between edge modes. 
By the Imry--Ma argument~\cite{imry1975:PhysRevLett.35.1399}, there is an optimal domain size $L$ that minimizes the total free energy density
\begin{align}
	f&=f_{0} + f_{\text{DW}} + f_{\text{dis}} + f_{\text{ent}} \nonumber\\
 &= f_{0} + 2 c (\Delta\phi)^2 \frac{1}{L^2} - 2\left(\frac{h\Delta\phi }{2} + \frac{ \pi T^2 }{6 v_{a}} + \frac{ \pi T^2 }{6 v_{b}}   \right) \frac{1}{L},
\end{align}
where $f_{0}$ is the domain energy of the corresponding phases. 
We have ignored the entropy of rearranging domains since it is small compared with the $2 c (\Delta\phi)^2$ term.
The free energy density is minimized when
\begin{equation}
	L=L^{*} = 2 c (\Delta\phi)^2 \left(\frac{h\Delta\phi }{2} + \frac{ \pi T^2 }{6 v_{a}} + \frac{ \pi T^2 }{6 v_{b}}   \right)^{-1},
 \label{eq:domain_size}
\end{equation}
and the minimized free energy density is 
\begin{equation}
	f=f_{0} - \frac{1}{2 c (\Delta\phi)} \left(\frac{h\Delta\phi }{2} + \frac{ \pi T^2 }{6 v_{a}} + \frac{ \pi T^2 }{6 v_{b}} \right)^2 .
\end{equation}
In order to reach the minimal free energy density, we have assumed that the total system size is
large enough for forming domains of size $L^{*}$. Eq.~(\ref{eq:domain_size}) implies that within our approximation of smooth domain walls, domains will be formed as long as $h\neq0$ or $T\neq0$.

Now we consider domain-dominated phases.
We assume a single edge mode for the uniform FCI phase.
We label the EQAH phase to be formed by domains of $-\phi_{\text{EQAH}}$ and $\phi_{\text{EQAH}}$, $f_{0}=\mathcal{F}(\varphi_{+})$, and its edge mode velocity is $v_{\text{EQAH}}$.
FCI phase is formed by domains of $-\phi_{\text{FCI}}$ and $\phi_{\text{FCI}}$, $f_{0}=\mathcal{F}(\varphi_{-})$, and its edge mode velocity is $v_{\text{FCI}}<v_{\text{EQAH}}$, as discussed in Sec.~\ref{sec2}. 
A mixed EQAH/FCI$+$ phase is formed by domains of $\phi_{\text{FCI}}$ and $\phi_{\text{EQAH}}$. 
Another mixed EQAH/FCI$-$ phase is formed by domains of $-\phi_{\text{FCI}}$ and $\phi_{\text{EQAH}}$.
$f_{0}$ for the two mixed phases is $[\mathcal{F}(\varphi_{+})+\mathcal{F}(\varphi_{-})]/2$.

\begin{figure}[t]
	\centering
		\includegraphics[width=\columnwidth]{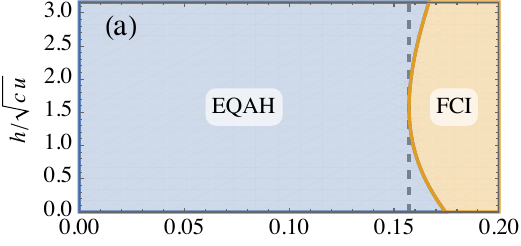}%
	
		\includegraphics[width=\columnwidth]{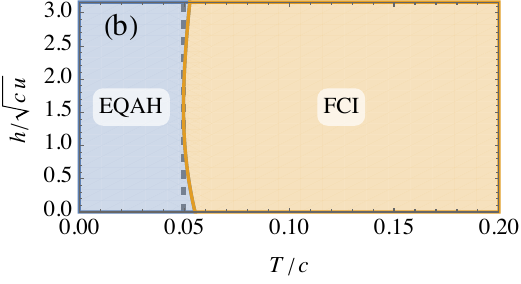}%
	\caption{Phase diagram of the competing EQAH and FCI phases when (a) $v_{\text{EQAH}}=0.1\sqrt{c^3/u}$ and $v_{\text{FCI}}= 0.03\sqrt{c^3/u}$; (b) $v_{\text{EQAH}}=0.01 \sqrt{c^3/u}$ and $v_{\text{FCI}}= 0.003 \sqrt{c^3/u}$. The other parameters are $s=0.5u$, $r=3u$, and $\Phi_{0}=1.3$. The dashed lines show the critical temperature $T_{1} = (3/\pi)^{1/2}  [\delta_{\varphi} \delta_{F} / ( \delta_{v}^{2}+ \delta_{\varphi}\delta_{v\varphi} )]^{1/4}$. The phases with domain walls between EQAH and FCI are not considered here. The phases shown in the figures represent domains with opposite Hall conductance.
    }
    \label{fig:phase_diag_smooth}
\end{figure}

{First, as a warm-up exercise, we focus on the competition between the phase with only EQAH domains and the phase with only FCI domains. This simplified scenario enables us to present analytical calculations in a concise manner. Later we will consider the domain walls between EQAH and FCI.}
The condition under which the FCI phase is favored over the EQAH phase is
\begin{equation}
	\delta_{\varphi} \left(\frac{h}{2}\right)^2 - 2 t^2 \delta_{v} \frac{h }{2} - t^4 \delta_{v\varphi} + \delta_{F} <0,
\end{equation}
where we have used the following parameters to simplify the equation:
\begin{align}
	\delta_{\varphi} &= \phi_{\rm EQAH}-\phi_{\rm FCI}, \\
	\delta_{v} &= \frac{1}{v_{\text{FCI}}}-\frac{1}{v_{\text{EQAH}}}, \\
	\delta_{v\varphi} &= \frac{1}{v_{\text{FCI}}^{2}\phi_{\rm FCI}} - \frac{1}{v_{\text{EQAH}}^{2}\phi_{\rm EQAH}}, \\
	\delta_{F} &= 2c [ \mathcal{F}(\varphi_{-}) - \mathcal{F}(\varphi_{+}) ], \\
	t^2 &= \frac{\pi T^2}{3}.
\end{align}
When $t<t_{1} =  [\delta_{\varphi} \delta_{F} / ( \delta_{v}^{2}+ \delta_{\varphi}\delta_{v\varphi} )]^{1/4}$,
the equation has no solution, which means the EQAH phase is always favored. 
When $t \geq t_{1} $, the solution is $ h_{-} < h < h_{+}$, where the two branches are given by
\begin{equation}
	\frac{h_{\pm} }{2} = \frac{t^2 \delta_{v} \pm \sqrt{t^4 (\delta_{v}^{2}+  \delta_{\varphi}\delta_{v\varphi})-\delta_{\varphi}\delta_{F}} } {\delta_{\varphi}}.
\end{equation}
$h_{+}$ is an increasing function of $t$, while $h_{-}$ is a decreasing function of $t$. 
They intersect at $h^{*} = 2 [ \delta_{F} \delta_{v}^{2} / \delta_{\varphi}( \delta_{v}^{2}+ \delta_{\varphi}\delta_{v\varphi} )]^{1/2} $ when $t=t_{1}$.
We also notice that $h_{-}<0$ when $t>t_{2} = (\delta_{F}/\delta_{v\varphi})^{1/4}$, suggesting that a clean (non-disordered) system starts to favor the FCI phase when $t\geq t_{2}$. 
A phase diagram showing the competition between the EQAH phase and the FCI phase is shown in Fig.~\ref{fig:phase_diag_smooth}, in which the two critical temperatures $t_{1}$ and $t_{2}$ are visible.

Now we consider the effect of interactions between the edge modes. {The edge modes in domain walls are co-propagating in both cases.}
We assume that the edge mode entropy of the EQAH and FCI phase are increased by a common factor $\Lambda$.
Since the entropy difference between the EQAH and FCI phase is increased, we expect that the FCI phase would be favored at a lower temperature. Increasing the edge entropy is equivalent to scaling the velocities $v_{\text{EQAH}} \to v_{\text{EQAH}} / \Lambda$ and $v_{\text{FCI}} \to v_{\text{FCI}} / \Lambda$ in the model above. As the result, parameters $\delta_{v}$ and $\delta_{v\varphi}$ would change to $\Lambda\delta_{v}$ and $\Lambda^2 \delta_{v\varphi}$. For the phase diagram, we have $t_{1,2}\to \Lambda^{-1/2} t_{1,2} $ and $h_{\pm} \sim \Lambda h_{\pm}$, but $h^{*}$ is invariant. Hence, the region of the EQAH phase shrinks in the phase diagram, agreeing with our expectation.

\begin{figure}[t]
	\centering
		\includegraphics[width=\columnwidth]{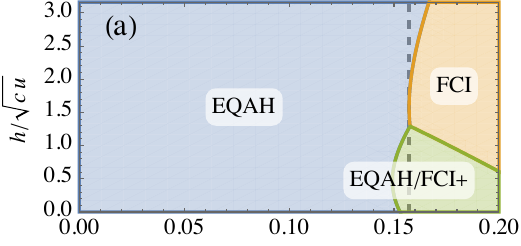}%
	
		\includegraphics[width=\columnwidth]{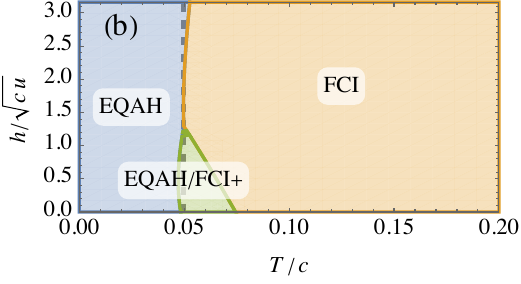}%
	\caption{Full phase diagram of four competing phases: EQAH, FCI, EQAH/FCI$+$, and EQAH/FCI$-$. (a) $v_{\text{EQAH}}=0.1\sqrt{c^3/u}$ and $v_{\text{FCI}}= 0.03\sqrt{c^3/u}$; (b) $v_{\text{EQAH}}=0.01 \sqrt{c^3/u}$ and $v_{\text{FCI}}= 0.003 \sqrt{c^3/u}$. We assume that there are no interactions between edge modes. The other parameters are $s=0.5u$, $r=3u$, $\Phi_{0}=1.3$. The dashed lines show the critical temperature $T_{1} = (3\hbar/\pi)^{1/2}  [\delta_{\varphi} \delta_{F} / ( \delta_{v}^{2}+ \delta_{\varphi}\delta_{v\varphi} )]^{1/4}$.
	}
	\label{fig:phase_diag_smooth_mix}
\end{figure}

Finally, we consider the other two possible phases, EQAH/FCI$+$, and EQAH/FCI$-$. 
Domain walls in EQAH/FCI$+$ have counter-propagating edge modes with velocities $v_{\text{EQAH}}$ and $v_{\text{FCI}}$, while domain walls in EQAH/FCI$-$ have co-propagating edge modes with velocities $v_{\text{EQAH}}$ and $v_{\text{FCI}}$.
We plot the phase diagram in Fig.~\ref{fig:phase_diag_smooth_mix} in the absence of edge mode interactions. It can be seen that when the disorder effect is weak, an intermediate EQAH/FCI$+$ phase would emerge during the phase transition from EQAH to FCI when the temperature increases.
We also notice that the EQAH/FCI$-$ phase is absent from this phase diagram, which can be explained by its higher energy cost to form domain walls.

{Due to the disorder and edge entropy, rich phases can emerge as shown in Figs. \ref{fig:phase_diag_smooth} and \ref{fig:phase_diag_smooth_mix}. Although FCI has higher energy compared to EQAH, it can be stabilized at higher temperature as a result of the higher edge entropy. Disorder can also cause the transition between EQAH and FCI. It is interesting to note that intermediate phases can appear with coexisting domains of EQAH and FCI. The Hall conductance of this intermediate phase depends on the population of EQAH and FCI, and their distribution in the devices.} {Under a training field, the quantized Hall conductance plateau is fully developed once the edge channel associated with the domains favored by the training field percolates the whole system.} {This could also explain the smoothness of the $T$-induced transition between the EQAH and FCI phases \cite{Lu_Han_Yao_Yang_Seo_Shi_Ye_Watanabe_Taniguchi_Ju_2024, Sarma_Xie_2024}: at an intermediate $T$, domains of EQAH and FCI can co-exist, with the population of the former smoothly decreasing as $T$ is increased.}

\section{Discussion and summary}\label{sec6}
Edge modes are natural contributors for the observed current- and particularly temperature-driven transition from EQAH to FCI, since they host gapless excitations while the bulk is gapped. In EQAH, the breaking of translational symmetry allows for the emergence of Goldstone modes, i.e., the phonon modes associated with the electron crystal. It is likely that these phonon modes are gapped due to the pinning of the crystal by impurities or hBN moir\'e potential as required by the quantization of the observed Hall conductance. If the phonon gap is much smaller than the EQAH and FCI gap, phonons of the electron crystal can contribute to the entropy and make the EQAH more favorable at higher temperature, which is inconsistent with the experiments.

The edge modes can be imaged using various experimental techniques that are already available. The population of the edges can be controlled by an external magnetic field since the domains sandwiching edges have opposite (fractional) Hall conductance, and hence opposite orbital magnetization. Disorder plays an important role in determining the domain populations and can also trigger transitions between EQAH and FCI. By keeping track of the transition temperature and threshold current versus domain wall population, the edge picture proposed here can be verified. We remark that full quantization of Hall conductance in transport measurement does not imply a single domain in the device. Instead, a full quantization of Hall conductance is already expected if one dominant domain percolates the device.

Theoretically, the current picture can be further substantiated by computing the ground-state energy of EQAH and FCI, and also edge modes in these two states using appropriate models for pentalayer graphene.
We hope that the current work can help motivate effort in this direction. We remark that the edge entropy mechanism is very general and should be applicable to various topological systems with competition between distinct topological phases.

To summarize, we proposed a scenario to explain the experimentally observed transition from EQAH to FCI either driven by temperature or current, based on the topologically protected gapless edge modes present in both phases. Our model is based on the assumption that the edge velocity in the FCI phase is lower than in the EQAH phase, resulting in higher entropy. Additionally, we show that disorder and temperature in the devices create domains with opposite fractionally quantized Hall conductance, leading to a network of edge modes. The velocity of these edge modes between domains is further reduced due to edge reconstruction. Current can also slow the edge velocity as the edge mode occupation is pushed near the gap edge. Consequently, the increase in edge entropy drives the transition from EQAH to FCI, induced either by temperature or current.

\begin{acknowledgements}
SZL would like to thank Long Ju for helpful discussion and for sharing the data prior to the publication. The work was carried out under the auspices of the U.S. DOE NNSA under Contract No. 89233218CNA000001 through the LDRD Program, and was performed, in part, at the Center for Integrated Nanotechnologies, an Office of Science User Facility operated for the U.S. DOE Office of Science, under user proposals $\#2018BU0010$ and $\#2018BU0083$. SZL acknowledges the hospitality of the Aspen Center for Physics (partially funded by the National Science Foundation) where the current work was motivated during a 2024 summer program.

\end{acknowledgements}

\emph{Note added---} During the preparation of the manuscript, we became aware of a related work, Ref.~\cite{Shavit_2024}, which also discusses the edge entropy driven transition between EQAH and FCI. No explanation of current driven transition was given in Ref.~\cite{Shavit_2024}. 

\appendix 
\section{Gap of FCI at different fillings}\label{appA}

\begin{figure}[t]
\begin{center}
\includegraphics[width = \columnwidth]{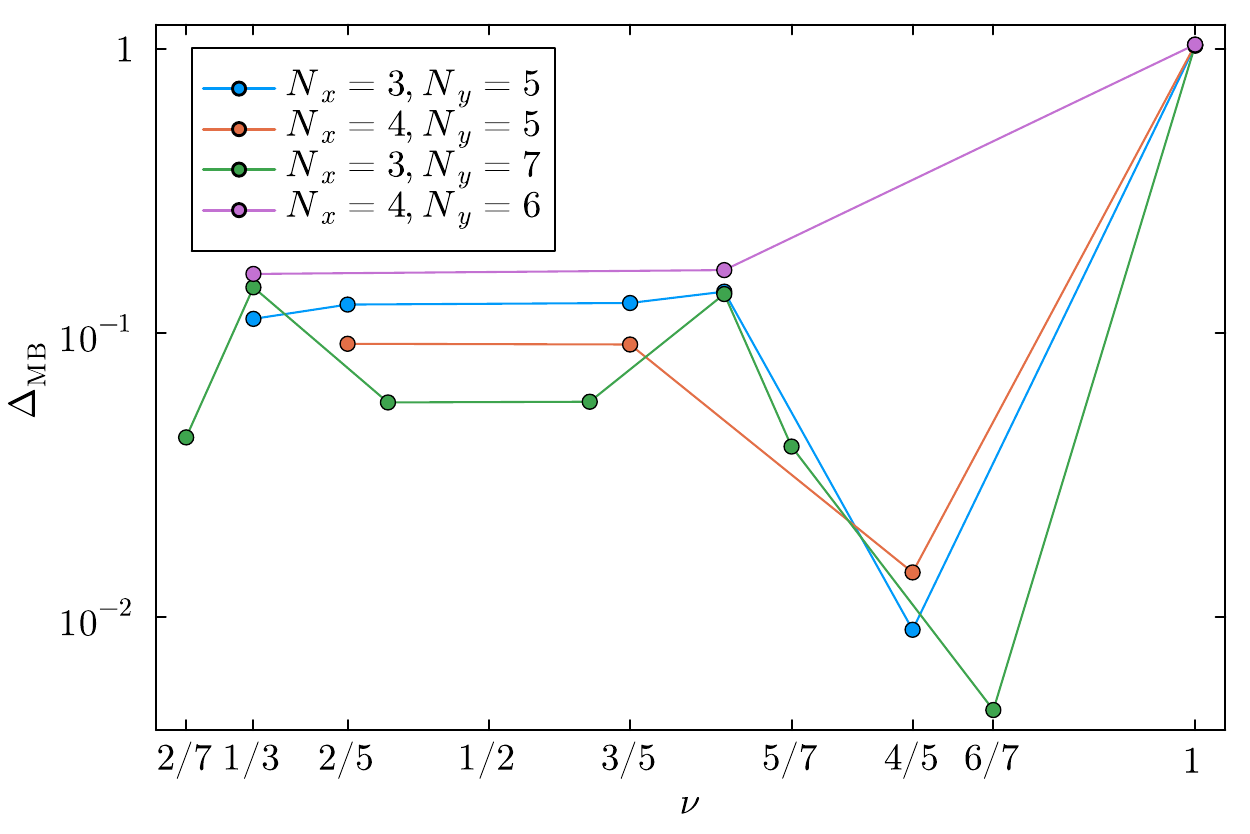}
\caption{Many-body gap $\Delta_{\rm MB}$ as a function of fillings $\nu$. $\Delta_{\rm MB}$ is defined as the gap in the many-body spectrum between degenerate FCI states and next excited state. %
}
\label{fig:MBgap}
\end{center}
\end{figure}
To illustrate the dependence of FCI gap on filling, we consider topological flatbands from the quadratic band crossing point with periodic strain introduced in Ref.~\cite{PhysRevLett.130.216401} as a model system. The chiral limit of this model gives topological flat bands with Chern number $C=\pm 1$, where the flatbands satisfy the trace condition. We focus on the second magic parameter, where the Berry curvature is very uniform. We then project the Coulomb interaction to the two middle flatbands and perform exact diagonalization; see Ref.~\cite{wu2024quantum} for details. FCI at different fractional fillings is stabilized due to the similarity between the lowest Landau level and the flat bands.

The results of the many-body gaps $\Delta_\mathrm{MB}$ in unit Coulomb energy $U=1$ as a function of the electron fillings $\nu$ are shown in Fig.~\ref{fig:MBgap}. One notable feature of $\Delta_\mathrm{MB}$ is that the Chern insulator at the integer filling has a much larger gap than the fractional fillings. The gap of FCI at fractional fillings is similar, except for the expected fragile FCI at $\nu=4/5$ and $\nu=6/7$. The finite-size scaling is different for different fractional fillings. For $\nu=n/3$, $\Delta_\mathrm{MB}$ saturates to some larger value as the size of the system increases, while for $\nu=n/5,n/7$,  $\Delta_\mathrm{MB}$ saturates to some lower values.

\section{Current-dependent edge entropy}\label{appB}

\begin{figure*}[t]

\centering
\includegraphics[width=1.5\columnwidth]{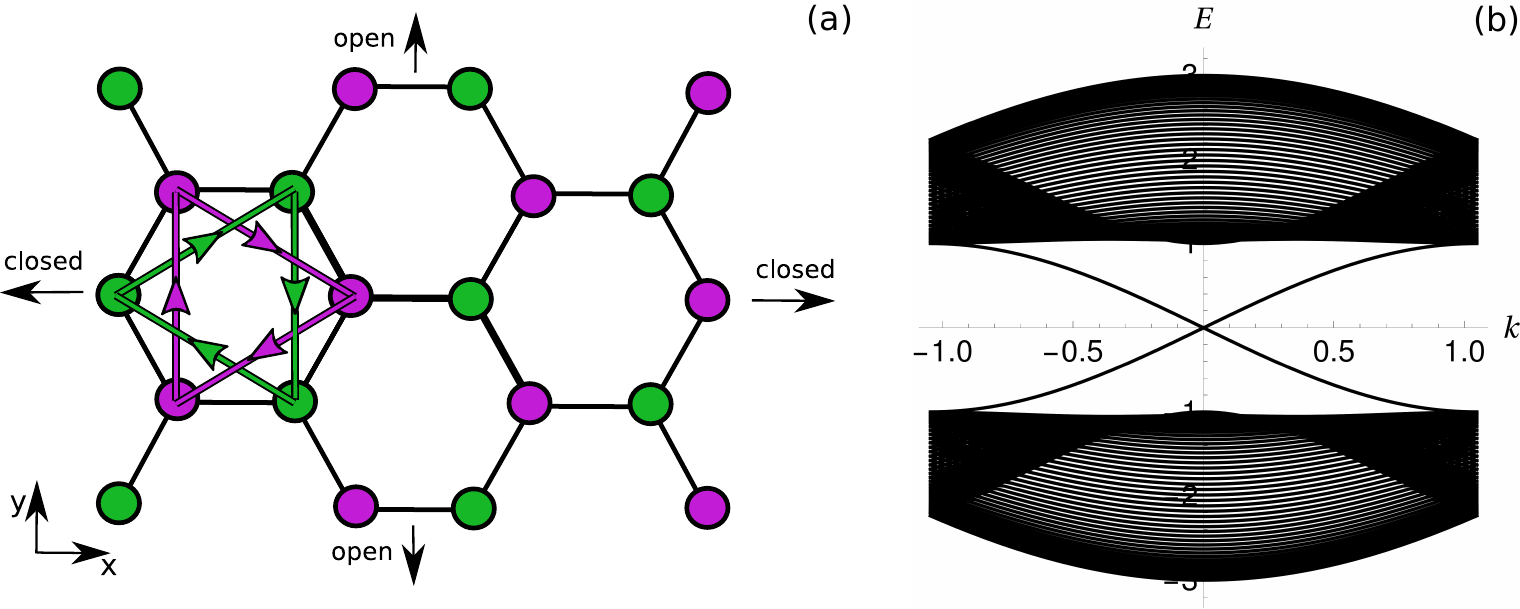}\caption{(a) Honeycomb lattice with armchair edges and illustration of Haldane
fluxes. (b) Energy spectrum for the Haldane model with armchair edges, for $t_{2}=0.2$
and $L=N_{k}=50$. \label{fig:geometry_armchair}}
\end{figure*}

To illustrate the current-dependent edge entropy, we consider the paradigmatic Haldane model \cite{PhysRevLett.61.2015}, where the entropy as a function of current can be calculated explicitly. This serves as a starting point to understand the edge entropy in EQAH and FCI. The Hamiltonian for the Haldane model is
\begin{equation}
H=-t\sum_{\langle i,j\rangle}c_{i}^{\dagger}c_{j}-t_{2}\sum_{\langle\langle i,j\rangle\rangle}e^{{\rm i}\phi_{ij}}c_{i}^{\dagger}c_{j}\,.
\end{equation}
The first term describes nearest-neighbor hoppings on the honeycomb
lattice. The second term describes next-nearest-neighbor hoppings,
where $\phi_{ij}=\pm\pi/2$ and the sign is defined by the arrows
depicted in Fig.~\ref{fig:geometry_armchair}: $+$ ($-$) if
the electron hops along (against) the direction of the arrow.

We compute the contribution to the thermal entropy
from the edge modes, and for that purpose, we consider a system
with open boundary conditions along the armchair edges, as illustrated
in Fig.~\ref{fig:geometry_armchair}(a).

In our calculations, we take $a=1$ as the length of the bond and the lattice
vectors ${\bf a}_{1}=(3,\sqrt{3})/2$ and ${\bf a}_{2}=(3,-\sqrt{3})/2$.
Writing
\[
c_{i}=\begin{cases}
a_{i}, & i\in A\\
b_{i}, & i\in B
\end{cases}\,,
\]
where $A$ and $B$ are the sublattices denoted in purple and green
respectively in Fig.~\ref{fig:geometry_armchair}(a), and performing
the Fourier transform along $x$ direction, we get
\begin{widetext}
\begin{align}
H(k_{x})={} & -t\sum_{y}a_{y,k_{x}}^{\dagger}(e^{-ik_{x}}b_{y,k_{x}}+e^{i\frac{1}{2}k{}_{x}}b_{y-1,k_{x}}+e^{i\frac{1}{2}k{}_{x}}b_{y+1,k_{x}})+\textrm{h.c.}\nonumber\\
 & -t_{2}\sum_{y}a_{y,k_{x}}^{\dagger}\Big(e^{-i\phi}a_{y+2,k_{x}}+e^{i\phi}a_{y-2,k_{x}}+2e^{i\phi}\cos\big(\frac{3}{2}k{}_{x}\big)a_{y+1,k_{x}}+2e^{-i\phi}\cos\big(\frac{3}{2}k{}_{x}\big)a_{y-1,k_{x}}\Big)+\textrm{h.c.}\nonumber\\
 & -t_{2}\sum_{y}b_{y,k_{x}}^{\dagger}\Big(e^{i\phi}b_{y+2,k_{x}}+e^{-i\phi}b_{y-2,k_{x}}+2e^{-i\phi}\cos\big(\frac{3}{2}k{}_{x}\big)b_{y+1,k_{x}}+2e^{i\phi}\cos\big(\frac{3}{2}k{}_{x}\big)b_{y-1,k_{x}}\Big)+\textrm{h.c.}
\end{align}
In order to single-out the entropy due to edge modes, we also compute
the bulk entropy. To do so, we use the Bloch Hamiltonian,
\begin{align}
H(k_{x},k_{y})={} & -t(e^{-ik_{x}}+e^{i(\frac{1}{2}k{}_{x}-\frac{\sqrt{3}}{2}k_{y})}+e^{i(\frac{1}{2}k{}_{x}+\frac{\sqrt{3}}{2}k_{y})})a_{k_{x}k_{y}}^{\dagger}b_{k_{x}k_{y}}+\textrm{h.c.}\nonumber\\
 & -2t_{2}[\cos(\phi-\frac{3}{2}k{}_{x}-\frac{\sqrt{3}}{2}k_{y})+\cos(\phi+\sqrt{3}k_{y})+\cos(\phi+\frac{3}{2}k{}_{x}-\frac{\sqrt{3}}{2}k_{y})]a_{k_{x}k_{y}}^{\dagger}a_{k_{x}k_{y}}+\textrm{h.c.}\nonumber\\
 & -2t_{2}[\cos(\phi+\frac{3}{2}k{}_{x}+\frac{\sqrt{3}}{2}k_{y})+\cos(\phi-\sqrt{3}k_{y})+\cos(\phi-\frac{3}{2}k{}_{x}+\frac{\sqrt{3}}{2}k_{y})]b_{k_{x}k_{y}}^{\dagger}b_{k_{x}k_{y}}+\textrm{h.c.}
\end{align}
\end{widetext}

By writing the Hamiltonian in the eigenbasis as $H=\sum_{\alpha}\epsilon_{\alpha}c_{\alpha}^{\dagger}c_{\alpha}$,
the thermal entropy can be easily computed through 
\begin{equation}
S=k_{B}\sum_{\alpha}\log(1+e^{-\beta(\epsilon_{\alpha}-\mu)})+\frac{1}{T}\sum_{\alpha}\frac{(\epsilon_{\alpha}-\mu)}{1+e^{\beta(\epsilon_{\alpha}-\mu)}}\,,
\end{equation}
where $\beta=1/(k_{B}T)$ and $\mu$ is the
chemical potential.

For the calculations in the case with open boundary conditions, we
will consider systems with $L$ unit cells along $y$ direction, and
a momentum grid with $N_{k}$ points. For the bulk calculation, we
will discretize the Brillouin zone in $N_{x}\times N_{y}$ points,
with $N_{x}=N_{k}$ and $N_{y}=L$. We also measure energy in units of the nearest-neighbor hopping strength
$t$.

We first reproduce the results for the energy spectrum with open boundary
conditions in Fig.~\ref{fig:geometry_armchair}(b). In this example,
it can be clearly seen that the velocity of the edge modes decreases
as they approach the gap edges, providing a concrete model example for the physical argument in Fig.~\ref{f1}(b).

\begin{figure*}[t]
\begin{centering}
\includegraphics[width=1.5\columnwidth]{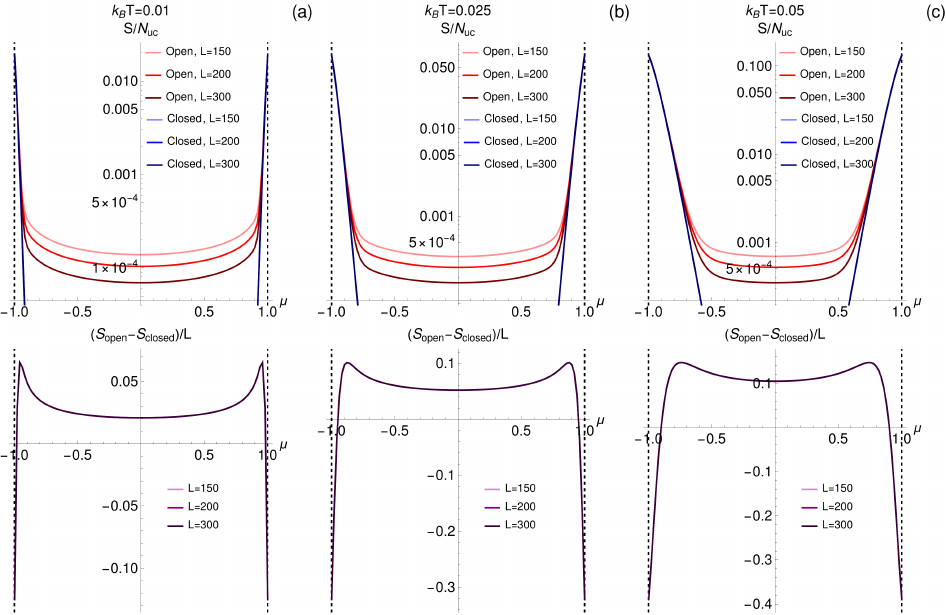}
\par\end{centering}
\caption{Thermal entropy due to bulk and edge with $t_{2}=0.2,N_{k}=L$, for
$T=0.01$ (a) $T=0.025$ (b) and $T=0.05$ (c). In the top panels
we compare the entropy computed for the closed and open systems normalized
to the number of unit cells $N_{uc}=LN_{k}$. The entropy due to edge
modes scales as $S\sim L$ and therefore we observe $S/N_{uc}\sim L^{-1}$
inside the energy gap for the open system. The bottom panel shows
the difference between the entropies of open and closed systems (normalized
to $L$), corresponding to the entropy contribution due to the edge
modes. The vertical dashed lines denote the bulk gap edges. \label{fig:entropy_bulk_edge}}
\end{figure*}

As depicted in Fig.~\ref{f1} (b), the current can be introduced by shifting $\mu$ in the opposite direction for the two counter-propagating edges. Therefore, we compute the thermal entropy for the open and closed system in Fig.~\ref{fig:entropy_bulk_edge}
as a function of the chemical potential $\mu$, and single out the
entropy due to the edge by subtracting the results using the set of parameters in Fig.~\ref{fig:geometry_armchair}(b). We can see that there
is an increase in edge entropy as $|\mu|$ departs from $\mu=0$,
which correlates well with the decrease in edge mode velocity. Furthermore,
the relative entropy gain becomes larger for smaller temperatures
which means that the contribution of the $-TS$
term to the free energy may be significant even for smaller temperatures.

For a sufficiently large $|\mu|$ (of the order of the gap), there is a sharp decrease
in the entropy difference. This occurs when the bulk states start contributing. Considering systems with the same volume,
there has to be a depletion of bulk states in the open system compared
to the closed system in order to accommodate for the edge states.
Because of this, the entropy of the open system is expected to decrease
with respect to the closed system when the bulk states start to contribute:
the open system has fewer bulk states and therefore smaller entropy. In experiments, we expect the current-driven transition between EQAH and FCI to occur before $|\mu|$ reaching the bulk because of the good quantization of the Hall conductivity.

\section{Domain configurations when the nonlinear term is dominant}\label{appC}

\subsection{Energy cost of a domain wall}
{In the main text, we considered the case where the potential well for the ordered phases in the free energy functional is shallow, so that order parameter can vary smoothly across domains. In this section, we consider the regime in which the potential well for order parameter is deep (or non-linear terms in the free energy dominate over the stiffness term), and the domain wall is sharp.}
To characterize a sharp domain wall separating phases of order parameters $\phi_{a}$ and $\phi_{b}$, we consider the following order parameter profile:
\begin{equation}
	\phi(\mathbf{x}) \equiv \phi(x,y) = \bar{\phi} + \Delta\phi \tanh\left(\frac{x}{\xi}\right),
\end{equation}
where $\bar{\phi}=(\phi_{a}+\phi_{b})/2$, $\Delta\phi=(\phi_{b}-\phi_{a})/2$, and $\xi$ is the typical width of the domain wall.
Compared with the case of the shallow potential well in the main text, calculations of the free energy cost of a domain wall are much more involved when the domain wall is sharp. 
The free energy can be written as
\begin{equation}
	F = \int d^2 \mathbf{x} \, \left[\mathcal{F}(\phi^2 -\Phi_0) + \frac{c}{2} (\nabla \phi)^2 \right],
\end{equation}
where $\mathcal{F}(\varphi) = - \frac{r}{2} \varphi^2 - s \varphi^3 + u \varphi^4$ is the auxiliary function we introduced in the main text.
We first focus on the non-linear term $\mathcal{F}(\phi^2 -\Phi_0)$ and consider the following expression:
\begin{widetext}
\begin{equation}
	\mathcal{F}(\phi^2 -\Phi_0) - \mathcal{F}(\tilde{\varphi}) - 	\tilde{\mathcal{F}} 
	= \sum_{k=1}^{4} a_{k} [(\bar{\phi})^2 + 2 \bar{\phi} \Delta\phi \tanh									(\frac{x}{\xi}) + (\Delta\phi)^2 \tanh^2(\frac{x}{\xi}) -\Phi_0 - \tilde{\varphi}]^k - 	\tilde{\mathcal{F}},
\end{equation}
where $\tilde{\varphi}$ and $\tilde{\mathcal{F}}$ are counterterms introduced for removing the divergence of the integral, and the coefficients $a_{k}$ are
\begin{align}
	a_{4} = u, \quad a_{3} = -s+4u \tilde{\varphi}, \quad a_{2} = - \frac{r}{2} - 3s \tilde{\varphi} + 6 u \tilde{\varphi}^2,
	\quad a_{1} = -r \tilde{\varphi} - 3s \tilde{\varphi}^2 + 4 u\tilde{\varphi}^3.
\end{align}
If we choose $\tilde{\varphi} = (\bar{\phi})^2 + (\Delta\phi)^2 - \Phi_{0}$ such that $(\bar{\phi})^2 - \Phi_{0} - \tilde{\varphi} = - (\Delta\phi)^2$, then
\begin{align}
		\mathcal{F}(\phi^2 -\Phi_0) - \mathcal{F}(\tilde{\varphi}) - 	\tilde{\mathcal{F}} 
		&= \sum_{k=1}^{4} \sum_{l=0}^{k} a_{k} \binom{k}{l} \left[ 2 \bar{\phi} \Delta\phi \tanh(\frac{x}{\xi}) \right]^{l} \left[ -  \frac{(\Delta\phi)^2}{\cosh^2(\frac{x}{\xi})} \right]^{k-l} - 	\tilde{\mathcal{F}}.
\end{align}
Since the terms containing an odd power of $\tanh(x/\xi)$ are odd and vanish under integration, we only keep terms with an even power of $\tanh(x/\xi)$,
\begin{align}
	\mathcal{F}(\phi^2 -\Phi_0) &- \mathcal{F}(\tilde{\varphi}) - 	\tilde{\mathcal{F}} 
	\sim \sum_{k=1}^{4} a_{k}  \left[ -  \frac{(\Delta\phi)^2}{\cosh^2(\frac{x}{\xi})} \right]^{k} + a_{4} \binom{4}{2} \left[ 2 \bar{\phi} \Delta\phi \tanh(\frac{x}{\xi}) \right]^{2} \left[ -  \frac{(\Delta\phi)^2}{\cosh^2(\frac{x}{\xi})} \right]^2\nonumber\\
	&+ a_{3} \binom{3}{2} \left[ 2 \bar{\phi} \Delta\phi \tanh(\frac{x}{\xi}) \right]^{2} \left[ -  \frac{(\Delta\phi)^2}{\cosh^2(\frac{x}{\xi})} \right] 
	+ a_{2} \left[ 2 \bar{\phi} \Delta\phi \tanh(\frac{x}{\xi}) \right]^{2} + a_{4} \left[ 2 \bar{\phi} \Delta\phi \tanh(\frac{x}{\xi}) \right]^{4} - 	\tilde{\mathcal{F}}.
\end{align}
We then choose $\tilde{\mathcal{F}}= a_{2} (2 \bar{\phi} \Delta\phi)^2 +  a_{4} (2\bar{\phi} \Delta\phi)^4$ such that
\begin{align}
	a_{2} \left[ 2 \bar{\phi} \Delta\phi \tanh(\frac{x}{\xi}) \right]^{2} + a_{4} \left[ 2 \bar{\phi} \Delta\phi \tanh(\frac{x}{\xi}) \right]^{4} - 	\tilde{\mathcal{F}} %
	={} a_4  (2 \bar{\phi} \Delta\phi)^4 \left[ - \frac{1}{\cosh^2(\frac{x}{\xi})}\right]^2
	+ [ a_{2} (2 \bar{\phi} \Delta\phi)^2 + 2 a_{4} (2\bar{\phi} \Delta\phi)^4 ] \left[- \frac{1}{\cosh^2(\frac{x}{\xi})} \right].
\end{align}
Finally, we can perform the integral
\begin{align}
	&\int_{-\infty}^{\infty} dx \, \mathcal{F}(\phi^2 -\Phi_0) - \mathcal{F}(\tilde{\varphi}) - 	\tilde{\mathcal{F}} = \xi g,
\end{align}
where
\begin{align}
	g ={}&  \sum_{k=1}^{4} \frac{2^{2k+1}}{2k} {}_2 F_{1} (k,2k;1+k;-1) a_{k}  \left[ -  (\Delta\phi)^2 \right]^{k} 
	+  \frac{4}{15} a_{4} \binom{4}{2} \left( 2 \bar{\phi} \Delta\phi  \right)^{2} \left[ -  (\Delta\phi)^2 \right]^2 \nonumber\\
	&+  \frac{2}{3} a_{3} \binom{3}{2} \left( 2 \bar{\phi} \Delta\phi  \right)^{2} \left[ -  (\Delta\phi)^2 \right] +  \frac{4}{3} a_4  (2 \bar{\phi} \Delta\phi)^4 - 2[ a_{2} (2 \bar{\phi} \Delta\phi)^2 + 2 a_{4} (2\bar{\phi} \Delta\phi)^4 ],
\end{align}
and ${}_2 F_{1}(a,b;c;z)$ is the hypergeometric function.
\end{widetext}

The integral of the stiffness term is easy to compute, and the result is
\begin{align}
	\int d x \, \frac{c}{2}  (\nabla \phi)^2 
	= \frac{c}{2} (\Delta\phi)^2 \int d x \, \frac{1}{\xi^2} \frac{1}{\cosh^4(\frac{x}{\xi})}
	= \frac{2c}{3\xi}  (\Delta\phi)^2 .
\end{align}
The free energy for a system of linear size $L$ is
\begin{equation}
	F = [\mathcal{F}(\tilde{\varphi})+	\tilde{\mathcal{F}}] L^2 +    \left( \xi g + \frac{1}{\xi} \frac{2c (\Delta\phi)^2}{3} \right) L.
\end{equation}
The energy is minimized at $\xi=\xi^{*} = \Delta\phi \sqrt{2c/3g}$, and the minimum energy is
\begin{equation}
	F = [\mathcal{F}(\tilde{\varphi})+	\tilde{\mathcal{F}}]  L^2 +  2 \Delta\phi \sqrt{2cg/3}  L.
\end{equation}
Therefore, $\mathcal{F}(\tilde{\varphi})+	\tilde{\mathcal{F}}$ is the free energy density in the bulk, and $2 \Delta\phi \sqrt{2cg/3}$ is the energy cost of a domain wall per unit length.

\subsection{Finite-temperature free energy density under disorder}
For the effect of the disorder, we consider a more sophisticated disorder field ${b}$ in this Appendix. It couples to the order parameter in the same way
\begin{align}
	F_{\text{dis}} = \int d^{2}\mathbf{x} \, b(\mathbf{x}) \phi(\mathbf{x}),
\end{align}
but its variance is Gaussian-correlated,
\begin{equation}
	\overline{b(\mathbf{x}) b(\mathbf{x}')} = \frac{h^2}{2\pi} \exp\left[ - \frac{(\mathbf{x}-\mathbf{x}')^2}{2 \lambda^2} \right],
\end{equation}
where $h$ is the disorder strength and $\lambda$ is the correlation length.
Then within a domain of linear length $L$ and order parameter $\phi$, the root mean square energy is~\cite{thomson2021:PhysRevB.103.125138}
\begin{equation}
    F_{\text{rms}} \approx \phi h \lambda L \sqrt{1- e^{-L^2/2\lambda^2}}.
\end{equation}
Using the same argument in Sec.~\ref{sec5}, for a domain-dominated phase, if the two possible order parameters are $\phi_{a}$ and $\phi_{b}$, the average energy density gain due to the disorder is
\begin{equation}
    f_{\text{dis}} \approx - \Delta\phi h \lambda \sqrt{1- e^{-L^2/2\lambda^2}} \frac{1}{L},
\end{equation}
where $\Delta\phi = (\phi_{b}-\phi_{a})/2$.
The analysis of edge state entropy is identical to that in Sec.~\ref{sec5}.

The finite-temperature free energy density of a uniform EQAH phase under disorder is still $f=\mathcal{F}(\varphi_{+})$. 
For a domain-dominated phase of $\phi_{a}$ and $\phi_{b}$, if each phase represented by a uniform order parameter $\phi_{a,b}$ has a single edge mode with velocity $v_{a,b}$, then the free energy density is
\begin{align}
	f ={} & \mathcal{F}(\tilde{\varphi})+\tilde{\mathcal{F}} + 4 \Delta \phi \sqrt{\frac{2 c g}{3}} \frac{1}{L}
	- h \lambda \Delta \phi \sqrt{1- e^{-L^2/2\lambda^2}} \frac{1}{L} \nonumber\\
 &- \frac{ \pi T^2 }{3}\left( \frac{1}{v_{a}}+\frac{1}{v_{b}}\right) \frac{1}{L} - T s_{0} \frac{1}{L^2},
\end{align}
where $s_{0}$ is a small domain rearrangement entropy per domain. 
We can ignore the term $Ts_{0}$ if either $s_{0}$ is small or the domain size $L $ is much greater than the thermal length $ v/T $.
Since the energy cost and gain are all of the same order $\mathcal{O}(L^{-1})$, one cannot immediately get the domain size from the free energy density, and the domain-dominated phase is not always favored over the uniform phase in the present case.
For a relatively strong disorder field, the typical domain size is close to the disorder correlation length $\lambda$~\cite{thomson2021:PhysRevB.103.125138}, and hence we choose $L \approxeq \lambda$.
{We also demand $L\gg \xi^{*}$, which is required by the basic assumptions of this Appendix.}

\subsection{Phase diagram}

\begin{figure*}[t]
	\centering
		\includegraphics[width=\columnwidth]{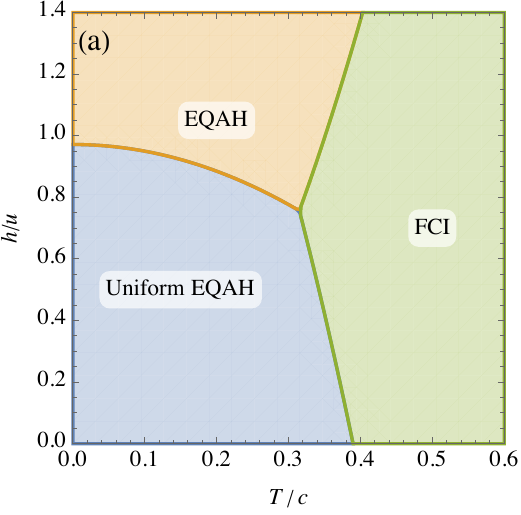}%
		\includegraphics[width=\columnwidth]{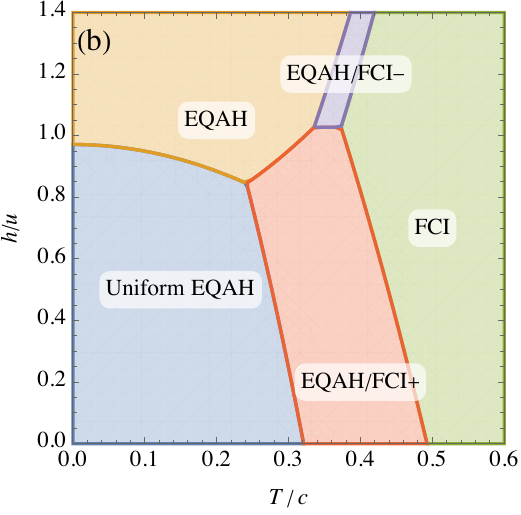}%
	\caption{Phase diagram of the competing phases when $v_{\text{FCI}}= 0.03 \sqrt{c^3/u}$ and $v_{\text{EQAH}}= 0.1 \sqrt{c^3/u}$. The other parameters are $s=0.5u$, $r=3u$, $\Phi_{0}=1.3$ and $L=\lambda=10\sqrt{c/u}$. If not explicitly mentioned as uniform phases, the phases shown in the figures represent domains with opposite Hall conductance. Domain wall width $\xi^{*}$ for the domain-dominated phases are: $0.674 \sqrt{c/u}$ for EQAH, $0.439\sqrt{c/u}$ for FCI, $0.326 \sqrt{c/u}$ for EQAH/FCI$+$, and $0.577\sqrt{c/u}$ for EQAH/FCI$-$.
	}
	\label{fig:phase_diag_sharp}
\end{figure*}

We start with three phases: a uniform EQAH phase, a domain-dominated EQAH phase, and a domain-dominated FCI phase.
The phase boundary between the uniform EQAH phase and the domain-dominated EQAH phase (we will refer to it as boundary I) is a parabola with a negative quadratic coefficient in the $h$--$T$ plane,
\begin{equation}
	 \frac{ 2\pi T^2 }{3v_{\text{EQAH}}} =
  \phi_{\text{EQAH}} \left(4\sqrt{\frac{2 c g_{\text{EQAH}}}{3}} - \alpha h \lambda  \right) ,
\end{equation}
where $\alpha=\sqrt{1- e^{-1/2}}$.
The domain-dominated EQAH phase is favored in the upper region of the parabola.
The phase boundary between the uniform EQAH phase and the domain-dominated FCI phase (boundary II) is also a parabola with a negative quadratic coefficient,
\begin{equation}
	 \frac{ 2\pi T^2}{3v_{\text{FCI}}}  = 
  \lambda\Delta \mathcal{F}   + \phi_{\text{FCI}} \left(4\sqrt{\frac{2 c g_{\text{FCI}}}{3}} - \alpha h \lambda  \right),
\end{equation}
where $\Delta \mathcal{F} = \mathcal{F}(\varphi_{-})-\mathcal{F}(\varphi_{+}) >0$.
Again, the domain-dominated FCI phase is favored in the upper region.
Since $\phi_{\text{EQAH}} > \phi_{\text{FCI}}$, the phase boundary between domain-dominated EQAH and FCI phases (boundary III) is a parabola with a positive quadratic coefficient,
\begin{align}
    \frac{ 2\pi T^2 }{3 v_{\text{FCI}}} -  \frac{ 2\pi T^2}{3 v_{\text{EQAH}}}
    ={}& \lambda\Delta \mathcal{F} 
    + \phi_{\text{FCI}} \left(4\sqrt{\frac{2 c g_{ \text{FCI}}}{3}} - \alpha h \lambda  \right) \nonumber\\
    &- \phi_{\text{EQAH}} \left(4\sqrt{\frac{2 c g_{\text{EQAH}}}{3}} - \alpha h \lambda \right).
\end{align}
The domain-dominated FCI phase is favored in the lower region of the parabola. {We also note that $g_{\text{FCI}} < g_{\text{EQAH}}$}.

To construct a phase diagram, we need the intersection point of the phase boundaries with the $h=0$ and $T=0$ lines. We use $T^*$ and $h^*$ to denote the corresponding $T$-intercept and $h$-intercept.
First, we compare $h^*$ for the three phase boundaries,
\begin{align}
    \alpha \lambda h^*_{ \text{I}} ={}& 4 \sqrt{\frac{2c g_{\rm EQAH}}{3}} \\
    \alpha \lambda h^*_{ \text{II}} ={}& 4\sqrt{\frac{2 c g_{\rm FCI}}{3}} +  \frac{\lambda\Delta \mathcal{F} }{\phi_{\rm FCI}}  \\
    \alpha \lambda h^*_{ \text{III}} ={}& 4 \sqrt{\frac{2c}{3}}\frac{\sqrt{g_{\rm FCI}} \phi_{\rm FCI} - \sqrt{g_{\rm EQAH}} \phi_{\rm EQAH}}{\phi_{\rm FCI}-\phi_{\rm EQAH}} \nonumber\\
    &+  \frac{\lambda\Delta \mathcal{F} }{\phi_{\rm FCI}-\phi_{\rm EQAH}} 
\end{align}
The relation between $h_{\text{II}}^{*}$ and $h_{\text{I}}^{*}$ is undetermined, so we 
discuss the two possibilities separately.

\paragraph*{(a) $h_{\mathrm{II}}^{*} < h_{\mathrm{I}}^{*} $. }
We notice the following identity:
\begin{equation}
	h^{*}_{\text{II}} - h^{*}_{\text{I}} = \frac{\phi_{\rm FCI}-\phi_{\rm EQAH}}{\phi_{\rm FCI}}  (h^{*}_{\text{III}} - h^{*}_{\text{I}}).
 \label{eq:h*_identity}
\end{equation}
As a result, $h_{\mathrm{II}}^{*}  < h_{\mathrm{I}}^{*} <  h_{\text{III}}^{*}$. 
Next, the intercepts $T^{*}$ for boundaries I and II are
\begin{align}
    \frac{ 2\pi (T_{\rm I}^*)^2}{3v_{\text{EQAH}}} &= 4\phi_{\text{EQAH}} \sqrt{\frac{2 c g_{\text{EQAH}}}{3}}  ,
	  \\
    \frac{2 \pi (T_{\rm II}^*)^2}{3v_{\text{FCI}}} &= \lambda\Delta \mathcal{F} + 4\phi_{\text{FCI}} \sqrt{\frac{2 c g_{\text{FCI}}}{3}} .
\end{align}
One can obtain the following identity:
\begin{align}
     \frac{ 2\pi (T_{\rm II}^*)^2}{3v_{\text{FCI}}} &-  \frac{ 2\pi (T_{\rm I}^*)^2}{3v_{\text{EQAH}}} = \phi_{\rm FCI} \alpha \lambda (h^*_{ \text{II}}-h^*_{ \text{I}}) \nonumber\\
     &+ 4(\phi_{\rm FCI}-\phi_{\rm EQAH}) \sqrt{\frac{2 c g_{\text{EQAH}}}{3}}.
     \label{eq:T*_identity}
\end{align}
Since both the left-hand and right-hand side of Eq.~(\ref{eq:T*_identity}) is negative, $T_{\rm II}^* < T_{\rm I}^*$, and hence the three phase boundaries would not intersect.

\paragraph*{(b) $h_{\mathrm{II}}^{*} > h_{\mathrm{I}}^{*} $.} 
By the same Eq.~(\ref{eq:h*_identity}), we get $h_{\text{III}}^{*}  < h_{\text{I}}^{*} <  h_{\text{II}}^{*}$ immediately.
From Eq.~(\ref{eq:T*_identity}), the relation between $T_{\rm II}^*$ and $ T_{\rm I}^*$ is still undetermined. However, since $h_{\text{III}}^{*}  < h_{\text{I}}^{*} <  h_{\text{II}}^{*}$ and the quadratic coefficient of boundary III is opposite to that of I and II, the three boundaries would intersect at a triple point. 
By sketching the boundaries, it can be seen that $T_{\rm II}^* < T_{\rm I}^*$ when $h_{\text{III}}^{*}>0$, and $h_{\text{III}}^{*}<0$ when $T_{\rm II}^* > T_{\rm I}^*$.

The phase diagram when $h_{\text{III}}^{*}  < h_{\text{I}}^{*} <  h_{\text{II}}^{*}$ and $T_{\rm II}^* < T_{\rm I}^*$ is shown in Fig.~\ref{fig:phase_diag_sharp}(a).
Comparing with the phase diagram in the smooth domain regime, Fig.~\ref{fig:phase_diag_smooth}, the high energy cost of a sharp domain wall would allow the presence of a uniform EQAH phase at a low temperature and weak disorder.

Taking the other two possible domain-dominated phases, i.e., EQAH/FCI$+$ and EQAH/FCI$-$, into consideration, the full phase diagram is shown in Fig.~\ref{fig:phase_diag_sharp}(b).
Both the EQAH/FCI$+$ and EQAH/FCI$-$ phase would become the intermediate phase at all three phase boundaries.
This is similar to what we got from the smooth domain wall results in Fig.~\ref{fig:phase_diag_smooth_mix}, expect for the presence of the uniform EQAH phase and the EQAH/FCI$-$ phase.
Again, we believe that the existence of the intermediate phases could explain the continuous temperature-induced transition observed in experiments.


\begin{thebibliography}{59}%
	\makeatletter
	\providecommand \@ifxundefined [1]{%
		\@ifx{#1\undefined}
	}%
	\providecommand \@ifnum [1]{%
		\ifnum #1\expandafter \@firstoftwo
		\else \expandafter \@secondoftwo
		\fi
	}%
	\providecommand \@ifx [1]{%
		\ifx #1\expandafter \@firstoftwo
		\else \expandafter \@secondoftwo
		\fi
	}%
	\providecommand \natexlab [1]{#1}%
	\providecommand \enquote  [1]{``#1''}%
	\providecommand \bibnamefont  [1]{#1}%
	\providecommand \bibfnamefont [1]{#1}%
	\providecommand \citenamefont [1]{#1}%
	\providecommand \href@noop [0]{\@secondoftwo}%
	\providecommand \href [0]{\begingroup \@sanitize@url \@href}%
	\providecommand \@href[1]{\@@startlink{#1}\@@href}%
	\providecommand \@@href[1]{\endgroup#1\@@endlink}%
	\providecommand \@sanitize@url [0]{\catcode `\\12\catcode `\$12\catcode `\&12\catcode `\#12\catcode `\^12\catcode `\_12\catcode `\%12\relax}%
	\providecommand \@@startlink[1]{}%
	\providecommand \@@endlink[0]{}%
	\providecommand \url  [0]{\begingroup\@sanitize@url \@url }%
	\providecommand \@url [1]{\endgroup\@href {#1}{\urlprefix }}%
	\providecommand \urlprefix  [0]{URL }%
	\providecommand \Eprint [0]{\href }%
	\providecommand \doibase [0]{https://doi.org/}%
	\providecommand \selectlanguage [0]{\@gobble}%
	\providecommand \bibinfo  [0]{\@secondoftwo}%
	\providecommand \bibfield  [0]{\@secondoftwo}%
	\providecommand \translation [1]{[#1]}%
	\providecommand \BibitemOpen [0]{}%
	\providecommand \bibitemStop [0]{}%
	\providecommand \bibitemNoStop [0]{.\EOS\space}%
	\providecommand \EOS [0]{\spacefactor3000\relax}%
	\providecommand \BibitemShut  [1]{\csname bibitem#1\endcsname}%
	\let\auto@bib@innerbib\@empty
	\bibitem [{\citenamefont {Tang}\ \emph {et~al.}(2011)\citenamefont {Tang}, \citenamefont {Mei},\ and\ \citenamefont {Wen}}]{PhysRevLett.106.236802}%
	\BibitemOpen
	\bibfield  {author} {\bibinfo {author} {\bibfnamefont {E.}~\bibnamefont {Tang}}, \bibinfo {author} {\bibfnamefont {J.-W.}\ \bibnamefont {Mei}},\ and\ \bibinfo {author} {\bibfnamefont {X.-G.}\ \bibnamefont {Wen}},\ }\bibfield  {title} {\bibinfo {title} {High-temperature fractional quantum {H}all states},\ }\href {https://doi.org/10.1103/PhysRevLett.106.236802} {\bibfield  {journal} {\bibinfo  {journal} {Phys. Rev. Lett.}\ }\textbf {\bibinfo {volume} {106}},\ \bibinfo {pages} {236802} (\bibinfo {year} {2011})}\BibitemShut {NoStop}%
	\bibitem [{\citenamefont {Sun}\ \emph {et~al.}(2011)\citenamefont {Sun}, \citenamefont {Gu}, \citenamefont {Katsura},\ and\ \citenamefont {Das~Sarma}}]{PhysRevLett.106.236803}%
	\BibitemOpen
	\bibfield  {author} {\bibinfo {author} {\bibfnamefont {K.}~\bibnamefont {Sun}}, \bibinfo {author} {\bibfnamefont {Z.}~\bibnamefont {Gu}}, \bibinfo {author} {\bibfnamefont {H.}~\bibnamefont {Katsura}},\ and\ \bibinfo {author} {\bibfnamefont {S.}~\bibnamefont {Das~Sarma}},\ }\bibfield  {title} {\bibinfo {title} {Nearly flatbands with nontrivial topology},\ }\href {https://doi.org/10.1103/PhysRevLett.106.236803} {\bibfield  {journal} {\bibinfo  {journal} {Phys. Rev. Lett.}\ }\textbf {\bibinfo {volume} {106}},\ \bibinfo {pages} {236803} (\bibinfo {year} {2011})}\BibitemShut {NoStop}%
	\bibitem [{\citenamefont {Neupert}\ \emph {et~al.}(2011)\citenamefont {Neupert}, \citenamefont {Santos}, \citenamefont {Chamon},\ and\ \citenamefont {Mudry}}]{PhysRevLett.106.236804}%
	\BibitemOpen
	\bibfield  {author} {\bibinfo {author} {\bibfnamefont {T.}~\bibnamefont {Neupert}}, \bibinfo {author} {\bibfnamefont {L.}~\bibnamefont {Santos}}, \bibinfo {author} {\bibfnamefont {C.}~\bibnamefont {Chamon}},\ and\ \bibinfo {author} {\bibfnamefont {C.}~\bibnamefont {Mudry}},\ }\bibfield  {title} {\bibinfo {title} {Fractional quantum {H}all states at zero magnetic field},\ }\href {https://doi.org/10.1103/PhysRevLett.106.236804} {\bibfield  {journal} {\bibinfo  {journal} {Phys. Rev. Lett.}\ }\textbf {\bibinfo {volume} {106}},\ \bibinfo {pages} {236804} (\bibinfo {year} {2011})}\BibitemShut {NoStop}%
	\bibitem [{\citenamefont {Sheng}\ \emph {et~al.}(2011)\citenamefont {Sheng}, \citenamefont {Gu}, \citenamefont {Sun},\ and\ \citenamefont {Sheng}}]{sheng2011fractional}%
	\BibitemOpen
	\bibfield  {author} {\bibinfo {author} {\bibfnamefont {D.~N.}\ \bibnamefont {Sheng}}, \bibinfo {author} {\bibfnamefont {Z.-C.}\ \bibnamefont {Gu}}, \bibinfo {author} {\bibfnamefont {K.}~\bibnamefont {Sun}},\ and\ \bibinfo {author} {\bibfnamefont {L.}~\bibnamefont {Sheng}},\ }\bibfield  {title} {\bibinfo {title} {Fractional quantum {H}all effect in the absence of {L}andau levels},\ }\href {https://doi.org/10.1038/ncomms1380} {\bibfield  {journal} {\bibinfo  {journal} {Nat. Commun.}\ }\textbf {\bibinfo {volume} {2}},\ \bibinfo {pages} {389} (\bibinfo {year} {2011})}\BibitemShut {NoStop}%
	\bibitem [{\citenamefont {Regnault}\ and\ \citenamefont {Bernevig}(2011)}]{PhysRevX.1.021014}%
	\BibitemOpen
	\bibfield  {author} {\bibinfo {author} {\bibfnamefont {N.}~\bibnamefont {Regnault}}\ and\ \bibinfo {author} {\bibfnamefont {B.~A.}\ \bibnamefont {Bernevig}},\ }\bibfield  {title} {\bibinfo {title} {Fractional {C}hern insulator},\ }\href {https://doi.org/10.1103/PhysRevX.1.021014} {\bibfield  {journal} {\bibinfo  {journal} {Phys. Rev. X}\ }\textbf {\bibinfo {volume} {1}},\ \bibinfo {pages} {021014} (\bibinfo {year} {2011})}\BibitemShut {NoStop}%
	\bibitem [{\citenamefont {Roy}(2014)}]{PhysRevB.90.165139}%
	\BibitemOpen
	\bibfield  {author} {\bibinfo {author} {\bibfnamefont {R.}~\bibnamefont {Roy}},\ }\bibfield  {title} {\bibinfo {title} {Band geometry of fractional topological insulators},\ }\href {https://doi.org/10.1103/PhysRevB.90.165139} {\bibfield  {journal} {\bibinfo  {journal} {Phys. Rev. B}\ }\textbf {\bibinfo {volume} {90}},\ \bibinfo {pages} {165139} (\bibinfo {year} {2014})}\BibitemShut {NoStop}%
	\bibitem [{\citenamefont {Ledwith}\ \emph {et~al.}(2020)\citenamefont {Ledwith}, \citenamefont {Tarnopolsky}, \citenamefont {Khalaf},\ and\ \citenamefont {Vishwanath}}]{PhysRevResearch.2.023237}%
	\BibitemOpen
	\bibfield  {author} {\bibinfo {author} {\bibfnamefont {P.~J.}\ \bibnamefont {Ledwith}}, \bibinfo {author} {\bibfnamefont {G.}~\bibnamefont {Tarnopolsky}}, \bibinfo {author} {\bibfnamefont {E.}~\bibnamefont {Khalaf}},\ and\ \bibinfo {author} {\bibfnamefont {A.}~\bibnamefont {Vishwanath}},\ }\bibfield  {title} {\bibinfo {title} {Fractional {C}hern insulator states in twisted bilayer graphene: An analytical approach},\ }\href {https://doi.org/10.1103/PhysRevResearch.2.023237} {\bibfield  {journal} {\bibinfo  {journal} {Phys. Rev. Res.}\ }\textbf {\bibinfo {volume} {2}},\ \bibinfo {pages} {023237} (\bibinfo {year} {2020})}\BibitemShut {NoStop}%
	\bibitem [{\citenamefont {Wang}\ \emph {et~al.}(2021)\citenamefont {Wang}, \citenamefont {Cano}, \citenamefont {Millis}, \citenamefont {Liu},\ and\ \citenamefont {Yang}}]{PhysRevLett.127.246403}%
	\BibitemOpen
	\bibfield  {author} {\bibinfo {author} {\bibfnamefont {J.}~\bibnamefont {Wang}}, \bibinfo {author} {\bibfnamefont {J.}~\bibnamefont {Cano}}, \bibinfo {author} {\bibfnamefont {A.~J.}\ \bibnamefont {Millis}}, \bibinfo {author} {\bibfnamefont {Z.}~\bibnamefont {Liu}},\ and\ \bibinfo {author} {\bibfnamefont {B.}~\bibnamefont {Yang}},\ }\bibfield  {title} {\bibinfo {title} {Exact {L}andau level description of geometry and interaction in a flatband},\ }\href {https://doi.org/10.1103/PhysRevLett.127.246403} {\bibfield  {journal} {\bibinfo  {journal} {Phys. Rev. Lett.}\ }\textbf {\bibinfo {volume} {127}},\ \bibinfo {pages} {246403} (\bibinfo {year} {2021})}\BibitemShut {NoStop}%
	\bibitem [{\citenamefont {Wang}\ \emph {et~al.}(2023)\citenamefont {Wang}, \citenamefont {Klevtsov},\ and\ \citenamefont {Liu}}]{PhysRevResearch.5.023167}%
	\BibitemOpen
	\bibfield  {author} {\bibinfo {author} {\bibfnamefont {J.}~\bibnamefont {Wang}}, \bibinfo {author} {\bibfnamefont {S.}~\bibnamefont {Klevtsov}},\ and\ \bibinfo {author} {\bibfnamefont {Z.}~\bibnamefont {Liu}},\ }\bibfield  {title} {\bibinfo {title} {Origin of model fractional {C}hern insulators in all topological ideal flatbands: Explicit color-entangled wave function and exact density algebra},\ }\href {https://doi.org/10.1103/PhysRevResearch.5.023167} {\bibfield  {journal} {\bibinfo  {journal} {Phys. Rev. Res.}\ }\textbf {\bibinfo {volume} {5}},\ \bibinfo {pages} {023167} (\bibinfo {year} {2023})}\BibitemShut {NoStop}%
	\bibitem [{\citenamefont {Ledwith}\ \emph {et~al.}(2023)\citenamefont {Ledwith}, \citenamefont {Vishwanath},\ and\ \citenamefont {Parker}}]{PhysRevB.108.205144}%
	\BibitemOpen
	\bibfield  {author} {\bibinfo {author} {\bibfnamefont {P.~J.}\ \bibnamefont {Ledwith}}, \bibinfo {author} {\bibfnamefont {A.}~\bibnamefont {Vishwanath}},\ and\ \bibinfo {author} {\bibfnamefont {D.~E.}\ \bibnamefont {Parker}},\ }\bibfield  {title} {\bibinfo {title} {Vortexability: A unifying criterion for ideal fractional {C}hern insulators},\ }\href {https://doi.org/10.1103/PhysRevB.108.205144} {\bibfield  {journal} {\bibinfo  {journal} {Phys. Rev. B}\ }\textbf {\bibinfo {volume} {108}},\ \bibinfo {pages} {205144} (\bibinfo {year} {2023})}\BibitemShut {NoStop}%
	\bibitem [{\citenamefont {Park}\ \emph {et~al.}(2023)\citenamefont {Park}, \citenamefont {Cai}, \citenamefont {Anderson}, \citenamefont {Zhang}, \citenamefont {Zhu}, \citenamefont {Liu}, \citenamefont {Wang}, \citenamefont {Holtzmann}, \citenamefont {Hu}, \citenamefont {Liu}, \citenamefont {Taniguchi}, \citenamefont {Watanabe}, \citenamefont {Chu}, \citenamefont {Cao}, \citenamefont {Fu}, \citenamefont {Yao}, \citenamefont {Chang}, \citenamefont {Cobden}, \citenamefont {Xiao},\ and\ \citenamefont {Xu}}]{Park2023}%
	\BibitemOpen
	\bibfield  {author} {\bibinfo {author} {\bibfnamefont {H.}~\bibnamefont {Park}}, \bibinfo {author} {\bibfnamefont {J.}~\bibnamefont {Cai}}, \bibinfo {author} {\bibfnamefont {E.}~\bibnamefont {Anderson}}, \bibinfo {author} {\bibfnamefont {Y.}~\bibnamefont {Zhang}}, \bibinfo {author} {\bibfnamefont {J.}~\bibnamefont {Zhu}}, \bibinfo {author} {\bibfnamefont {X.}~\bibnamefont {Liu}}, \bibinfo {author} {\bibfnamefont {C.}~\bibnamefont {Wang}}, \bibinfo {author} {\bibfnamefont {W.}~\bibnamefont {Holtzmann}}, \bibinfo {author} {\bibfnamefont {C.}~\bibnamefont {Hu}}, \bibinfo {author} {\bibfnamefont {Z.}~\bibnamefont {Liu}}, \bibinfo {author} {\bibfnamefont {T.}~\bibnamefont {Taniguchi}}, \bibinfo {author} {\bibfnamefont {K.}~\bibnamefont {Watanabe}}, \bibinfo {author} {\bibfnamefont {J.-H.}\ \bibnamefont {Chu}}, \bibinfo {author} {\bibfnamefont {T.}~\bibnamefont {Cao}}, \bibinfo {author} {\bibfnamefont {L.}~\bibnamefont {Fu}}, \bibinfo {author} {\bibfnamefont {W.}~\bibnamefont {Yao}}, \bibinfo {author}
		{\bibfnamefont {C.-Z.}\ \bibnamefont {Chang}}, \bibinfo {author} {\bibfnamefont {D.}~\bibnamefont {Cobden}}, \bibinfo {author} {\bibfnamefont {D.}~\bibnamefont {Xiao}},\ and\ \bibinfo {author} {\bibfnamefont {X.}~\bibnamefont {Xu}},\ }\bibfield  {title} {\bibinfo {title} {Observation of fractionally quantized anomalous {H}all effect},\ }\href {https://doi.org/10.1038/s41586-023-06536-0} {\bibfield  {journal} {\bibinfo  {journal} {Nature}\ }\textbf {\bibinfo {volume} {622}},\ \bibinfo {pages} {74} (\bibinfo {year} {2023})}\BibitemShut {NoStop}%
	\bibitem [{\citenamefont {Xu}\ \emph {et~al.}(2023)\citenamefont {Xu}, \citenamefont {Sun}, \citenamefont {Jia}, \citenamefont {Liu}, \citenamefont {Xu}, \citenamefont {Li}, \citenamefont {Gu}, \citenamefont {Watanabe}, \citenamefont {Taniguchi}, \citenamefont {Tong}, \citenamefont {Jia}, \citenamefont {Shi}, \citenamefont {Jiang}, \citenamefont {Zhang}, \citenamefont {Liu},\ and\ \citenamefont {Li}}]{PhysRevX.13.031037}%
	\BibitemOpen
	\bibfield  {author} {\bibinfo {author} {\bibfnamefont {F.}~\bibnamefont {Xu}}, \bibinfo {author} {\bibfnamefont {Z.}~\bibnamefont {Sun}}, \bibinfo {author} {\bibfnamefont {T.}~\bibnamefont {Jia}}, \bibinfo {author} {\bibfnamefont {C.}~\bibnamefont {Liu}}, \bibinfo {author} {\bibfnamefont {C.}~\bibnamefont {Xu}}, \bibinfo {author} {\bibfnamefont {C.}~\bibnamefont {Li}}, \bibinfo {author} {\bibfnamefont {Y.}~\bibnamefont {Gu}}, \bibinfo {author} {\bibfnamefont {K.}~\bibnamefont {Watanabe}}, \bibinfo {author} {\bibfnamefont {T.}~\bibnamefont {Taniguchi}}, \bibinfo {author} {\bibfnamefont {B.}~\bibnamefont {Tong}}, \bibinfo {author} {\bibfnamefont {J.}~\bibnamefont {Jia}}, \bibinfo {author} {\bibfnamefont {Z.}~\bibnamefont {Shi}}, \bibinfo {author} {\bibfnamefont {S.}~\bibnamefont {Jiang}}, \bibinfo {author} {\bibfnamefont {Y.}~\bibnamefont {Zhang}}, \bibinfo {author} {\bibfnamefont {X.}~\bibnamefont {Liu}},\ and\ \bibinfo {author} {\bibfnamefont {T.}~\bibnamefont {Li}},\ }\bibfield  {title} {\bibinfo {title}
		{Observation of integer and fractional quantum anomalous {H}all effects in twisted bilayer {MoTe}$_{2}$},\ }\href {https://doi.org/10.1103/PhysRevX.13.031037} {\bibfield  {journal} {\bibinfo  {journal} {Phys. Rev. X}\ }\textbf {\bibinfo {volume} {13}},\ \bibinfo {pages} {031037} (\bibinfo {year} {2023})}\BibitemShut {NoStop}%
	\bibitem [{\citenamefont {Cai}\ \emph {et~al.}(2023)\citenamefont {Cai}, \citenamefont {Anderson}, \citenamefont {Wang}, \citenamefont {Zhang}, \citenamefont {Liu}, \citenamefont {Holtzmann}, \citenamefont {Zhang}, \citenamefont {Fan}, \citenamefont {Taniguchi}, \citenamefont {Watanabe}, \citenamefont {Ran}, \citenamefont {Cao}, \citenamefont {Fu}, \citenamefont {Xiao}, \citenamefont {Yao},\ and\ \citenamefont {Xu}}]{Cai2023}%
	\BibitemOpen
	\bibfield  {author} {\bibinfo {author} {\bibfnamefont {J.}~\bibnamefont {Cai}}, \bibinfo {author} {\bibfnamefont {E.}~\bibnamefont {Anderson}}, \bibinfo {author} {\bibfnamefont {C.}~\bibnamefont {Wang}}, \bibinfo {author} {\bibfnamefont {X.}~\bibnamefont {Zhang}}, \bibinfo {author} {\bibfnamefont {X.}~\bibnamefont {Liu}}, \bibinfo {author} {\bibfnamefont {W.}~\bibnamefont {Holtzmann}}, \bibinfo {author} {\bibfnamefont {Y.}~\bibnamefont {Zhang}}, \bibinfo {author} {\bibfnamefont {F.}~\bibnamefont {Fan}}, \bibinfo {author} {\bibfnamefont {T.}~\bibnamefont {Taniguchi}}, \bibinfo {author} {\bibfnamefont {K.}~\bibnamefont {Watanabe}}, \bibinfo {author} {\bibfnamefont {Y.}~\bibnamefont {Ran}}, \bibinfo {author} {\bibfnamefont {T.}~\bibnamefont {Cao}}, \bibinfo {author} {\bibfnamefont {L.}~\bibnamefont {Fu}}, \bibinfo {author} {\bibfnamefont {D.}~\bibnamefont {Xiao}}, \bibinfo {author} {\bibfnamefont {W.}~\bibnamefont {Yao}},\ and\ \bibinfo {author} {\bibfnamefont {X.}~\bibnamefont {Xu}},\ }\bibfield  {title}
	{\bibinfo {title} {Signatures of fractional quantum anomalous {H}all states in twisted {MoTe}$_{2}$},\ }\href {https://doi.org/10.1038/s41586-023-06289-w} {\bibfield  {journal} {\bibinfo  {journal} {Nature}\ }\textbf {\bibinfo {volume} {622}},\ \bibinfo {pages} {63} (\bibinfo {year} {2023})}\BibitemShut {NoStop}%
	\bibitem [{\citenamefont {Zeng}\ \emph {et~al.}(2023)\citenamefont {Zeng}, \citenamefont {Xia}, \citenamefont {Kang}, \citenamefont {Zhu}, \citenamefont {Kn{\"u}ppel}, \citenamefont {Vaswani}, \citenamefont {Watanabe}, \citenamefont {Taniguchi}, \citenamefont {Mak},\ and\ \citenamefont {Shan}}]{Zeng2023}%
	\BibitemOpen
	\bibfield  {author} {\bibinfo {author} {\bibfnamefont {Y.}~\bibnamefont {Zeng}}, \bibinfo {author} {\bibfnamefont {Z.}~\bibnamefont {Xia}}, \bibinfo {author} {\bibfnamefont {K.}~\bibnamefont {Kang}}, \bibinfo {author} {\bibfnamefont {J.}~\bibnamefont {Zhu}}, \bibinfo {author} {\bibfnamefont {P.}~\bibnamefont {Kn{\"u}ppel}}, \bibinfo {author} {\bibfnamefont {C.}~\bibnamefont {Vaswani}}, \bibinfo {author} {\bibfnamefont {K.}~\bibnamefont {Watanabe}}, \bibinfo {author} {\bibfnamefont {T.}~\bibnamefont {Taniguchi}}, \bibinfo {author} {\bibfnamefont {K.~F.}\ \bibnamefont {Mak}},\ and\ \bibinfo {author} {\bibfnamefont {J.}~\bibnamefont {Shan}},\ }\bibfield  {title} {\bibinfo {title} {Thermodynamic evidence of fractional {C}hern insulator in moir{\'e} {MoTe}$_{2}$},\ }\href {https://doi.org/10.1038/s41586-023-06452-3} {\bibfield  {journal} {\bibinfo  {journal} {Nature}\ }\textbf {\bibinfo {volume} {622}},\ \bibinfo {pages} {69} (\bibinfo {year} {2023})}\BibitemShut {NoStop}%
	\bibitem [{\citenamefont {Li}\ \emph {et~al.}(2021)\citenamefont {Li}, \citenamefont {Kumar}, \citenamefont {Sun},\ and\ \citenamefont {Lin}}]{PhysRevResearch.3.L032070}%
	\BibitemOpen
	\bibfield  {author} {\bibinfo {author} {\bibfnamefont {H.}~\bibnamefont {Li}}, \bibinfo {author} {\bibfnamefont {U.}~\bibnamefont {Kumar}}, \bibinfo {author} {\bibfnamefont {K.}~\bibnamefont {Sun}},\ and\ \bibinfo {author} {\bibfnamefont {S.-Z.}\ \bibnamefont {Lin}},\ }\bibfield  {title} {\bibinfo {title} {Spontaneous fractional {C}hern insulators in transition metal dichalcogenide moir\'e superlattices},\ }\href {https://doi.org/10.1103/PhysRevResearch.3.L032070} {\bibfield  {journal} {\bibinfo  {journal} {Phys. Rev. Res.}\ }\textbf {\bibinfo {volume} {3}},\ \bibinfo {pages} {L032070} (\bibinfo {year} {2021})}\BibitemShut {NoStop}%
	\bibitem [{\citenamefont {Reddy}\ \emph {et~al.}(2023)\citenamefont {Reddy}, \citenamefont {Alsallom}, \citenamefont {Zhang}, \citenamefont {Devakul},\ and\ \citenamefont {Fu}}]{PhysRevB.108.085117}%
	\BibitemOpen
	\bibfield  {author} {\bibinfo {author} {\bibfnamefont {A.~P.}\ \bibnamefont {Reddy}}, \bibinfo {author} {\bibfnamefont {F.}~\bibnamefont {Alsallom}}, \bibinfo {author} {\bibfnamefont {Y.}~\bibnamefont {Zhang}}, \bibinfo {author} {\bibfnamefont {T.}~\bibnamefont {Devakul}},\ and\ \bibinfo {author} {\bibfnamefont {L.}~\bibnamefont {Fu}},\ }\bibfield  {title} {\bibinfo {title} {Fractional quantum anomalous {H}all states in twisted bilayer {MoTe}$_{2}$ and {WSe}$_{2}$},\ }\href {https://doi.org/10.1103/PhysRevB.108.085117} {\bibfield  {journal} {\bibinfo  {journal} {Phys. Rev. B}\ }\textbf {\bibinfo {volume} {108}},\ \bibinfo {pages} {085117} (\bibinfo {year} {2023})}\BibitemShut {NoStop}%
	\bibitem [{\citenamefont {Morales-Dur\'an}\ \emph {et~al.}(2024)\citenamefont {Morales-Dur\'an}, \citenamefont {Wei}, \citenamefont {Shi},\ and\ \citenamefont {MacDonald}}]{PhysRevLett.132.096602}%
	\BibitemOpen
	\bibfield  {author} {\bibinfo {author} {\bibfnamefont {N.}~\bibnamefont {Morales-Dur\'an}}, \bibinfo {author} {\bibfnamefont {N.}~\bibnamefont {Wei}}, \bibinfo {author} {\bibfnamefont {J.}~\bibnamefont {Shi}},\ and\ \bibinfo {author} {\bibfnamefont {A.~H.}\ \bibnamefont {MacDonald}},\ }\bibfield  {title} {\bibinfo {title} {Magic angles and fractional {C}hern insulators in twisted homobilayer transition metal dichalcogenides},\ }\href {https://doi.org/10.1103/PhysRevLett.132.096602} {\bibfield  {journal} {\bibinfo  {journal} {Phys. Rev. Lett.}\ }\textbf {\bibinfo {volume} {132}},\ \bibinfo {pages} {096602} (\bibinfo {year} {2024})}\BibitemShut {NoStop}%
	\bibitem [{\citenamefont {Lu}\ \emph {et~al.}(2024{\natexlab{a}})\citenamefont {Lu}, \citenamefont {Han}, \citenamefont {Yao}, \citenamefont {Reddy}, \citenamefont {Yang}, \citenamefont {Seo}, \citenamefont {Watanabe}, \citenamefont {Taniguchi}, \citenamefont {Fu},\ and\ \citenamefont {Ju}}]{Lu2024}%
	\BibitemOpen
	\bibfield  {author} {\bibinfo {author} {\bibfnamefont {Z.}~\bibnamefont {Lu}}, \bibinfo {author} {\bibfnamefont {T.}~\bibnamefont {Han}}, \bibinfo {author} {\bibfnamefont {Y.}~\bibnamefont {Yao}}, \bibinfo {author} {\bibfnamefont {A.~P.}\ \bibnamefont {Reddy}}, \bibinfo {author} {\bibfnamefont {J.}~\bibnamefont {Yang}}, \bibinfo {author} {\bibfnamefont {J.}~\bibnamefont {Seo}}, \bibinfo {author} {\bibfnamefont {K.}~\bibnamefont {Watanabe}}, \bibinfo {author} {\bibfnamefont {T.}~\bibnamefont {Taniguchi}}, \bibinfo {author} {\bibfnamefont {L.}~\bibnamefont {Fu}},\ and\ \bibinfo {author} {\bibfnamefont {L.}~\bibnamefont {Ju}},\ }\bibfield  {title} {\bibinfo {title} {Fractional quantum anomalous {H}all effect in multilayer graphene},\ }\href {https://doi.org/10.1038/s41586-023-07010-7} {\bibfield  {journal} {\bibinfo  {journal} {Nature}\ }\textbf {\bibinfo {volume} {626}},\ \bibinfo {pages} {759} (\bibinfo {year} {2024}{\natexlab{a}})}\BibitemShut {NoStop}%
	\bibitem [{\citenamefont {Lu}\ \emph {et~al.}(2025)\citenamefont {Lu}, \citenamefont {Han}, \citenamefont {Yao}, \citenamefont {Hadjri}, \citenamefont {Yang}, \citenamefont {Seo}, \citenamefont {Shi}, \citenamefont {Ye}, \citenamefont {Watanabe}, \citenamefont {Taniguchi},\ and\ \citenamefont {Ju}}]{Lu_Han_Yao_Yang_Seo_Shi_Ye_Watanabe_Taniguchi_Ju_2024}%
	\BibitemOpen
	\bibfield  {author} {\bibinfo {author} {\bibfnamefont {Z.}~\bibnamefont {Lu}}, \bibinfo {author} {\bibfnamefont {T.}~\bibnamefont {Han}}, \bibinfo {author} {\bibfnamefont {Y.}~\bibnamefont {Yao}}, \bibinfo {author} {\bibfnamefont {Z.}~\bibnamefont {Hadjri}}, \bibinfo {author} {\bibfnamefont {J.}~\bibnamefont {Yang}}, \bibinfo {author} {\bibfnamefont {J.}~\bibnamefont {Seo}}, \bibinfo {author} {\bibfnamefont {L.}~\bibnamefont {Shi}}, \bibinfo {author} {\bibfnamefont {S.}~\bibnamefont {Ye}}, \bibinfo {author} {\bibfnamefont {K.}~\bibnamefont {Watanabe}}, \bibinfo {author} {\bibfnamefont {T.}~\bibnamefont {Taniguchi}},\ and\ \bibinfo {author} {\bibfnamefont {L.}~\bibnamefont {Ju}},\ }\bibfield  {title} {\bibinfo {title} {Extended quantum anomalous {H}all states in graphene/h{BN} moir{\'e} superlattices},\ }\href {https://doi.org/10.1038/s41586-024-08470-1} {\bibfield  {journal} {\bibinfo  {journal} {Nature}\ }\textbf {\bibinfo {volume} {637}},\ \bibinfo {pages} {1090} (\bibinfo {year} {2025})}\BibitemShut
	{NoStop}%
	\bibitem [{\citenamefont {Xie}\ \emph {et~al.}(2024)\citenamefont {Xie}, \citenamefont {Huo}, \citenamefont {Lu}, \citenamefont {Feng}, \citenamefont {Zhang}, \citenamefont {Wang}, \citenamefont {Yang}, \citenamefont {Watanabe}, \citenamefont {Taniguchi}, \citenamefont {Liu}, \citenamefont {Song}, \citenamefont {Xie}, \citenamefont {Liu},\ and\ \citenamefont {Lu}}]{xie2024evenodddenominatorfractionalquantum}%
	\BibitemOpen
	\bibfield  {author} {\bibinfo {author} {\bibfnamefont {J.}~\bibnamefont {Xie}}, \bibinfo {author} {\bibfnamefont {Z.}~\bibnamefont {Huo}}, \bibinfo {author} {\bibfnamefont {X.}~\bibnamefont {Lu}}, \bibinfo {author} {\bibfnamefont {Z.}~\bibnamefont {Feng}}, \bibinfo {author} {\bibfnamefont {Z.}~\bibnamefont {Zhang}}, \bibinfo {author} {\bibfnamefont {W.}~\bibnamefont {Wang}}, \bibinfo {author} {\bibfnamefont {Q.}~\bibnamefont {Yang}}, \bibinfo {author} {\bibfnamefont {K.}~\bibnamefont {Watanabe}}, \bibinfo {author} {\bibfnamefont {T.}~\bibnamefont {Taniguchi}}, \bibinfo {author} {\bibfnamefont {K.}~\bibnamefont {Liu}}, \bibinfo {author} {\bibfnamefont {Z.}~\bibnamefont {Song}}, \bibinfo {author} {\bibfnamefont {X.~C.}\ \bibnamefont {Xie}}, \bibinfo {author} {\bibfnamefont {J.}~\bibnamefont {Liu}},\ and\ \bibinfo {author} {\bibfnamefont {X.}~\bibnamefont {Lu}},\ }\bibfield  {title} {\bibinfo {title} {Even- and odd-denominator fractional quantum anomalous {H}all effect in graphene moire superlattices},\
	}\Eprint {https://arxiv.org/abs/2405.16944} {arXiv:2405.16944 [cond-mat.mes-hall]}  (\bibinfo {year} {2024})\BibitemShut {NoStop}%
	\bibitem [{\citenamefont {Waters}\ \emph {et~al.}(2024)\citenamefont {Waters}, \citenamefont {Okounkova}, \citenamefont {Su}, \citenamefont {Zhou}, \citenamefont {Yao}, \citenamefont {Watanabe}, \citenamefont {Taniguchi}, \citenamefont {Xu}, \citenamefont {Zhang}, \citenamefont {Folk},\ and\ \citenamefont {Yankowitz}}]{Waters_Okounkova2024}%
	\BibitemOpen
	\bibfield  {author} {\bibinfo {author} {\bibfnamefont {D.}~\bibnamefont {Waters}}, \bibinfo {author} {\bibfnamefont {A.}~\bibnamefont {Okounkova}}, \bibinfo {author} {\bibfnamefont {R.}~\bibnamefont {Su}}, \bibinfo {author} {\bibfnamefont {B.}~\bibnamefont {Zhou}}, \bibinfo {author} {\bibfnamefont {J.}~\bibnamefont {Yao}}, \bibinfo {author} {\bibfnamefont {K.}~\bibnamefont {Watanabe}}, \bibinfo {author} {\bibfnamefont {T.}~\bibnamefont {Taniguchi}}, \bibinfo {author} {\bibfnamefont {X.}~\bibnamefont {Xu}}, \bibinfo {author} {\bibfnamefont {Y.-H.}\ \bibnamefont {Zhang}}, \bibinfo {author} {\bibfnamefont {J.}~\bibnamefont {Folk}},\ and\ \bibinfo {author} {\bibfnamefont {M.}~\bibnamefont {Yankowitz}},\ }\bibfield  {title} {\bibinfo {title} {Interplay of electronic crystals with integer and fractional {C}hern insulators in moir\'e pentalayer graphene},\ }\Eprint {https://arxiv.org/abs/2408.10133} {arXiv:2408.10133 [cond-mat.mes-hall]}  (\bibinfo {year} {2024})\BibitemShut {NoStop}%
	\bibitem [{\citenamefont {Wu}\ \emph {et~al.}(2019)\citenamefont {Wu}, \citenamefont {Lovorn}, \citenamefont {Tutuc}, \citenamefont {Martin},\ and\ \citenamefont {MacDonald}}]{PhysRevLett.122.086402}%
	\BibitemOpen
	\bibfield  {author} {\bibinfo {author} {\bibfnamefont {F.}~\bibnamefont {Wu}}, \bibinfo {author} {\bibfnamefont {T.}~\bibnamefont {Lovorn}}, \bibinfo {author} {\bibfnamefont {E.}~\bibnamefont {Tutuc}}, \bibinfo {author} {\bibfnamefont {I.}~\bibnamefont {Martin}},\ and\ \bibinfo {author} {\bibfnamefont {A.~H.}\ \bibnamefont {MacDonald}},\ }\bibfield  {title} {\bibinfo {title} {Topological insulators in twisted transition metal dichalcogenide homobilayers},\ }\href {https://doi.org/10.1103/PhysRevLett.122.086402} {\bibfield  {journal} {\bibinfo  {journal} {Phys. Rev. Lett.}\ }\textbf {\bibinfo {volume} {122}},\ \bibinfo {pages} {086402} (\bibinfo {year} {2019})}\BibitemShut {NoStop}%
	\bibitem [{\citenamefont {Li}\ \emph {et~al.}(2024)\citenamefont {Li}, \citenamefont {Su}, \citenamefont {Kim}, \citenamefont {Kee}, \citenamefont {Sun},\ and\ \citenamefont {Lin}}]{li2024contrasting}%
	\BibitemOpen
	\bibfield  {author} {\bibinfo {author} {\bibfnamefont {H.}~\bibnamefont {Li}}, \bibinfo {author} {\bibfnamefont {Y.}~\bibnamefont {Su}}, \bibinfo {author} {\bibfnamefont {Y.~B.}\ \bibnamefont {Kim}}, \bibinfo {author} {\bibfnamefont {H.-Y.}\ \bibnamefont {Kee}}, \bibinfo {author} {\bibfnamefont {K.}~\bibnamefont {Sun}},\ and\ \bibinfo {author} {\bibfnamefont {S.-Z.}\ \bibnamefont {Lin}},\ }\bibfield  {title} {\bibinfo {title} {Contrasting twisted bilayer graphene and transition metal dichalcogenides for fractional {C}hern insulators: An emergent gauge picture},\ }\href {https://doi.org/10.1103/PhysRevB.109.245131} {\bibfield  {journal} {\bibinfo  {journal} {Phys. Rev. B}\ }\textbf {\bibinfo {volume} {109}},\ \bibinfo {pages} {245131} (\bibinfo {year} {2024})}\BibitemShut {NoStop}%
	\bibitem [{\citenamefont {Zhou}\ \emph {et~al.}(2024)\citenamefont {Zhou}, \citenamefont {Yang},\ and\ \citenamefont {Zhang}}]{zhou2023fractionalquantumanomaloushall}%
	\BibitemOpen
	\bibfield  {author} {\bibinfo {author} {\bibfnamefont {B.}~\bibnamefont {Zhou}}, \bibinfo {author} {\bibfnamefont {H.}~\bibnamefont {Yang}},\ and\ \bibinfo {author} {\bibfnamefont {Y.-H.}\ \bibnamefont {Zhang}},\ }\bibfield  {title} {\bibinfo {title} {Fractional quantum anomalous {H}all effect in rhombohedral multilayer graphene in the moir\'eless limit},\ }\href {https://doi.org/10.1103/PhysRevLett.133.206504} {\bibfield  {journal} {\bibinfo  {journal} {Phys. Rev. Lett.}\ }\textbf {\bibinfo {volume} {133}},\ \bibinfo {pages} {206504} (\bibinfo {year} {2024})}\BibitemShut {NoStop}%
	\bibitem [{\citenamefont {Dong}\ \emph {et~al.}(2024{\natexlab{a}})\citenamefont {Dong}, \citenamefont {Wang}, \citenamefont {Wang}, \citenamefont {Soejima}, \citenamefont {Zaletel}, \citenamefont {Vishwanath},\ and\ \citenamefont {Parker}}]{dong2023anomaloushallcrystalsrhombohedral}%
	\BibitemOpen
	\bibfield  {author} {\bibinfo {author} {\bibfnamefont {J.}~\bibnamefont {Dong}}, \bibinfo {author} {\bibfnamefont {T.}~\bibnamefont {Wang}}, \bibinfo {author} {\bibfnamefont {T.}~\bibnamefont {Wang}}, \bibinfo {author} {\bibfnamefont {T.}~\bibnamefont {Soejima}}, \bibinfo {author} {\bibfnamefont {M.~P.}\ \bibnamefont {Zaletel}}, \bibinfo {author} {\bibfnamefont {A.}~\bibnamefont {Vishwanath}},\ and\ \bibinfo {author} {\bibfnamefont {D.~E.}\ \bibnamefont {Parker}},\ }\bibfield  {title} {\bibinfo {title} {Anomalous {H}all crystals in rhombohedral multilayer graphene. {I}. {I}nteraction-driven {C}hern bands and fractional quantum {H}all states at zero magnetic field},\ }\href {https://doi.org/10.1103/PhysRevLett.133.206503} {\bibfield  {journal} {\bibinfo  {journal} {Phys. Rev. Lett.}\ }\textbf {\bibinfo {volume} {133}},\ \bibinfo {pages} {206503} (\bibinfo {year} {2024}{\natexlab{a}})}\BibitemShut {NoStop}%
	\bibitem [{\citenamefont {Tan}\ and\ \citenamefont {Devakul}(2024)}]{tan2024parentberrycurvatureideal}%
	\BibitemOpen
	\bibfield  {author} {\bibinfo {author} {\bibfnamefont {T.}~\bibnamefont {Tan}}\ and\ \bibinfo {author} {\bibfnamefont {T.}~\bibnamefont {Devakul}},\ }\bibfield  {title} {\bibinfo {title} {Parent {B}erry curvature and the ideal anomalous {H}all crystal},\ }\href {https://doi.org/10.1103/PhysRevX.14.041040} {\bibfield  {journal} {\bibinfo  {journal} {Phys. Rev. X}\ }\textbf {\bibinfo {volume} {14}},\ \bibinfo {pages} {041040} (\bibinfo {year} {2024})}\BibitemShut {NoStop}%
	\bibitem [{\citenamefont {Dong}\ \emph {et~al.}(2024{\natexlab{b}})\citenamefont {Dong}, \citenamefont {Patri},\ and\ \citenamefont {Senthil}}]{Dong_Patri_Senthil_2024}%
	\BibitemOpen
	\bibfield  {author} {\bibinfo {author} {\bibfnamefont {Z.}~\bibnamefont {Dong}}, \bibinfo {author} {\bibfnamefont {A.~S.}\ \bibnamefont {Patri}},\ and\ \bibinfo {author} {\bibfnamefont {T.}~\bibnamefont {Senthil}},\ }\bibfield  {title} {\bibinfo {title} {Stability of anomalous {H}all crystals in multilayer rhombohedral graphene},\ }\href {https://doi.org/10.1103/PhysRevB.110.205130} {\bibfield  {journal} {\bibinfo  {journal} {Phys. Rev. B}\ }\textbf {\bibinfo {volume} {110}},\ \bibinfo {pages} {205130} (\bibinfo {year} {2024}{\natexlab{b}})}\BibitemShut {NoStop}%
	\bibitem [{\citenamefont {Soejima}\ \emph {et~al.}(2024)\citenamefont {Soejima}, \citenamefont {Dong}, \citenamefont {Wang}, \citenamefont {Wang}, \citenamefont {Zaletel}, \citenamefont {Vishwanath},\ and\ \citenamefont {Parker}}]{soejima2024anomaloushallcrystalsrhombohedral}%
	\BibitemOpen
	\bibfield  {author} {\bibinfo {author} {\bibfnamefont {T.}~\bibnamefont {Soejima}}, \bibinfo {author} {\bibfnamefont {J.}~\bibnamefont {Dong}}, \bibinfo {author} {\bibfnamefont {T.}~\bibnamefont {Wang}}, \bibinfo {author} {\bibfnamefont {T.}~\bibnamefont {Wang}}, \bibinfo {author} {\bibfnamefont {M.~P.}\ \bibnamefont {Zaletel}}, \bibinfo {author} {\bibfnamefont {A.}~\bibnamefont {Vishwanath}},\ and\ \bibinfo {author} {\bibfnamefont {D.~E.}\ \bibnamefont {Parker}},\ }\bibfield  {title} {\bibinfo {title} {Anomalous {H}all crystals in rhombohedral multilayer graphene. {II}. {G}eneral mechanism and a minimal model},\ }\href {https://doi.org/10.1103/PhysRevB.110.205124} {\bibfield  {journal} {\bibinfo  {journal} {Phys. Rev. B}\ }\textbf {\bibinfo {volume} {110}},\ \bibinfo {pages} {205124} (\bibinfo {year} {2024})}\BibitemShut {NoStop}%
	\bibitem [{\citenamefont {Herzog-Arbeitman}\ \emph {et~al.}(2024)\citenamefont {Herzog-Arbeitman}, \citenamefont {Wang}, \citenamefont {Liu}, \citenamefont {Tam}, \citenamefont {Qi}, \citenamefont {Jia}, \citenamefont {Efetov}, \citenamefont {Vafek}, \citenamefont {Regnault}, \citenamefont {Weng}, \citenamefont {Wu}, \citenamefont {Bernevig},\ and\ \citenamefont {Yu}}]{PhysRevB.109.205122}%
	\BibitemOpen
	\bibfield  {author} {\bibinfo {author} {\bibfnamefont {J.}~\bibnamefont {Herzog-Arbeitman}}, \bibinfo {author} {\bibfnamefont {Y.}~\bibnamefont {Wang}}, \bibinfo {author} {\bibfnamefont {J.}~\bibnamefont {Liu}}, \bibinfo {author} {\bibfnamefont {P.~M.}\ \bibnamefont {Tam}}, \bibinfo {author} {\bibfnamefont {Z.}~\bibnamefont {Qi}}, \bibinfo {author} {\bibfnamefont {Y.}~\bibnamefont {Jia}}, \bibinfo {author} {\bibfnamefont {D.~K.}\ \bibnamefont {Efetov}}, \bibinfo {author} {\bibfnamefont {O.}~\bibnamefont {Vafek}}, \bibinfo {author} {\bibfnamefont {N.}~\bibnamefont {Regnault}}, \bibinfo {author} {\bibfnamefont {H.}~\bibnamefont {Weng}}, \bibinfo {author} {\bibfnamefont {Q.}~\bibnamefont {Wu}}, \bibinfo {author} {\bibfnamefont {B.~A.}\ \bibnamefont {Bernevig}},\ and\ \bibinfo {author} {\bibfnamefont {J.}~\bibnamefont {Yu}},\ }\bibfield  {title} {\bibinfo {title} {Moir\'e fractional {C}hern insulators. {II}. {F}irst-principles calculations and continuum models of rhombohedral graphene superlattices},\ }\href
	{https://doi.org/10.1103/PhysRevB.109.205122} {\bibfield  {journal} {\bibinfo  {journal} {Phys. Rev. B}\ }\textbf {\bibinfo {volume} {109}},\ \bibinfo {pages} {205122} (\bibinfo {year} {2024})}\BibitemShut {NoStop}%
	\bibitem [{\citenamefont {Kwan}\ \emph {et~al.}(2023)\citenamefont {Kwan}, \citenamefont {Yu}, \citenamefont {Herzog-Arbeitman}, \citenamefont {Efetov}, \citenamefont {Regnault},\ and\ \citenamefont {Bernevig}}]{Kwan_Yu_Herzog-Arbeitman_Efetov_Regnault_Bernevig_2023}%
	\BibitemOpen
	\bibfield  {author} {\bibinfo {author} {\bibfnamefont {Y.~H.}\ \bibnamefont {Kwan}}, \bibinfo {author} {\bibfnamefont {J.}~\bibnamefont {Yu}}, \bibinfo {author} {\bibfnamefont {J.}~\bibnamefont {Herzog-Arbeitman}}, \bibinfo {author} {\bibfnamefont {D.~K.}\ \bibnamefont {Efetov}}, \bibinfo {author} {\bibfnamefont {N.}~\bibnamefont {Regnault}},\ and\ \bibinfo {author} {\bibfnamefont {B.~A.}\ \bibnamefont {Bernevig}},\ }\bibfield  {title} {\bibinfo {title} {Moir\'e fractional {C}hern insulators {III}: {H}artree-{F}ock phase diagram, magic angle regime for {C}hern insulator states, the role of the moir\'e potential and {G}oldstone gaps in rhombohedral graphene superlattices},\ }\Eprint {https://arxiv.org/abs/2312.11617} {arXiv:2312.11617 [cond-mat.str-el]}  (\bibinfo {year} {2023})\BibitemShut {NoStop}%
	\bibitem [{\citenamefont {Yu}\ \emph {et~al.}(2024)\citenamefont {Yu}, \citenamefont {Herzog-Arbeitman}, \citenamefont {Kwan}, \citenamefont {Regnault},\ and\ \citenamefont {Bernevig}}]{Yu_Herzog-Arbeitman_Kwan_Regnault_Bernevig_2024}%
	\BibitemOpen
	\bibfield  {author} {\bibinfo {author} {\bibfnamefont {J.}~\bibnamefont {Yu}}, \bibinfo {author} {\bibfnamefont {J.}~\bibnamefont {Herzog-Arbeitman}}, \bibinfo {author} {\bibfnamefont {Y.~H.}\ \bibnamefont {Kwan}}, \bibinfo {author} {\bibfnamefont {N.}~\bibnamefont {Regnault}},\ and\ \bibinfo {author} {\bibfnamefont {B.~A.}\ \bibnamefont {Bernevig}},\ }\bibfield  {title} {\bibinfo {title} {Moir\'e fractional {C}hern insulators {IV}: Fluctuation-driven collapse of {FCI}s in multi-band exact diagonalization calculations on rhombohedral graphene},\ }\Eprint {https://arxiv.org/abs/2407.13770} {arXiv:2407.13770 [cond-mat.str-el]}  (\bibinfo {year} {2024})\BibitemShut {NoStop}%
	\bibitem [{\citenamefont {Guo}\ \emph {et~al.}(2024)\citenamefont {Guo}, \citenamefont {Lu}, \citenamefont {Xie},\ and\ \citenamefont {Liu}}]{PhysRevB.110.075109}%
	\BibitemOpen
	\bibfield  {author} {\bibinfo {author} {\bibfnamefont {Z.}~\bibnamefont {Guo}}, \bibinfo {author} {\bibfnamefont {X.}~\bibnamefont {Lu}}, \bibinfo {author} {\bibfnamefont {B.}~\bibnamefont {Xie}},\ and\ \bibinfo {author} {\bibfnamefont {J.}~\bibnamefont {Liu}},\ }\bibfield  {title} {\bibinfo {title} {Fractional {C}hern insulator states in multilayer graphene moir\'e superlattices},\ }\href {https://doi.org/10.1103/PhysRevB.110.075109} {\bibfield  {journal} {\bibinfo  {journal} {Phys. Rev. B}\ }\textbf {\bibinfo {volume} {110}},\ \bibinfo {pages} {075109} (\bibinfo {year} {2024})}\BibitemShut {NoStop}%
	\bibitem [{\citenamefont {Das~Sarma}\ and\ \citenamefont {Xie}(2024)}]{Sarma_Xie_2024}%
	\BibitemOpen
	\bibfield  {author} {\bibinfo {author} {\bibfnamefont {S.}~\bibnamefont {Das~Sarma}}\ and\ \bibinfo {author} {\bibfnamefont {M.}~\bibnamefont {Xie}},\ }\bibfield  {title} {\bibinfo {title} {Thermal crossover from a {C}hern insulator to a fractional {C}hern insulator in pentalayer graphene},\ }\href {https://doi.org/10.1103/PhysRevB.110.155148} {\bibfield  {journal} {\bibinfo  {journal} {Phys. Rev. B}\ }\textbf {\bibinfo {volume} {110}},\ \bibinfo {pages} {155148} (\bibinfo {year} {2024})}\BibitemShut {NoStop}%
	\bibitem [{\citenamefont {Patri}\ \emph {et~al.}(2024)\citenamefont {Patri}, \citenamefont {Dong},\ and\ \citenamefont {Senthil}}]{Patri_Dong_Senthil_2024}%
	\BibitemOpen
	\bibfield  {author} {\bibinfo {author} {\bibfnamefont {A.~S.}\ \bibnamefont {Patri}}, \bibinfo {author} {\bibfnamefont {Z.}~\bibnamefont {Dong}},\ and\ \bibinfo {author} {\bibfnamefont {T.}~\bibnamefont {Senthil}},\ }\bibfield  {title} {\bibinfo {title} {Extended quantum anomalous {H}all effect in moir\'e structures: Phase transitions and transport},\ }\href {https://doi.org/10.1103/PhysRevB.110.245115} {\bibfield  {journal} {\bibinfo  {journal} {Phys. Rev. B}\ }\textbf {\bibinfo {volume} {110}},\ \bibinfo {pages} {245115} (\bibinfo {year} {2024})}\BibitemShut {NoStop}%
	\bibitem [{\citenamefont {Sheng}\ \emph {et~al.}(2024)\citenamefont {Sheng}, \citenamefont {Reddy}, \citenamefont {Abouelkomsan}, \citenamefont {Bergholtz},\ and\ \citenamefont {Fu}}]{sheng2024quantumanomaloushallcrystal}%
	\BibitemOpen
	\bibfield  {author} {\bibinfo {author} {\bibfnamefont {D.~N.}\ \bibnamefont {Sheng}}, \bibinfo {author} {\bibfnamefont {A.~P.}\ \bibnamefont {Reddy}}, \bibinfo {author} {\bibfnamefont {A.}~\bibnamefont {Abouelkomsan}}, \bibinfo {author} {\bibfnamefont {E.~J.}\ \bibnamefont {Bergholtz}},\ and\ \bibinfo {author} {\bibfnamefont {L.}~\bibnamefont {Fu}},\ }\bibfield  {title} {\bibinfo {title} {Quantum anomalous {H}all crystal at fractional filling of moir\'e superlattices},\ }\href {https://doi.org/10.1103/PhysRevLett.133.066601} {\bibfield  {journal} {\bibinfo  {journal} {Phys. Rev. Lett.}\ }\textbf {\bibinfo {volume} {133}},\ \bibinfo {pages} {066601} (\bibinfo {year} {2024})}\BibitemShut {NoStop}%
	\bibitem [{\citenamefont {Zeng}\ \emph {et~al.}(2024)\citenamefont {Zeng}, \citenamefont {Guerci}, \citenamefont {Cr\'epel}, \citenamefont {Millis},\ and\ \citenamefont {Cano}}]{PhysRevLett.132.236601}%
	\BibitemOpen
	\bibfield  {author} {\bibinfo {author} {\bibfnamefont {Y.}~\bibnamefont {Zeng}}, \bibinfo {author} {\bibfnamefont {D.}~\bibnamefont {Guerci}}, \bibinfo {author} {\bibfnamefont {V.}~\bibnamefont {Cr\'epel}}, \bibinfo {author} {\bibfnamefont {A.~J.}\ \bibnamefont {Millis}},\ and\ \bibinfo {author} {\bibfnamefont {J.}~\bibnamefont {Cano}},\ }\bibfield  {title} {\bibinfo {title} {Sublattice structure and topology in spontaneously crystallized electronic states},\ }\href {https://doi.org/10.1103/PhysRevLett.132.236601} {\bibfield  {journal} {\bibinfo  {journal} {Phys. Rev. Lett.}\ }\textbf {\bibinfo {volume} {132}},\ \bibinfo {pages} {236601} (\bibinfo {year} {2024})}\BibitemShut {NoStop}%
	\bibitem [{\citenamefont {Lu}\ and\ \citenamefont {Ran}(2012)}]{PhysRevB.85.165134}%
	\BibitemOpen
	\bibfield  {author} {\bibinfo {author} {\bibfnamefont {Y.-M.}\ \bibnamefont {Lu}}\ and\ \bibinfo {author} {\bibfnamefont {Y.}~\bibnamefont {Ran}},\ }\bibfield  {title} {\bibinfo {title} {Symmetry-protected fractional {C}hern insulators and fractional topological insulators},\ }\href {https://doi.org/10.1103/PhysRevB.85.165134} {\bibfield  {journal} {\bibinfo  {journal} {Phys. Rev. B}\ }\textbf {\bibinfo {volume} {85}},\ \bibinfo {pages} {165134} (\bibinfo {year} {2012})}\BibitemShut {NoStop}%
	\bibitem [{\citenamefont {Giamarchi}(2003)}]{giamarchi2003quantum}%
	\BibitemOpen
	\bibfield  {author} {\bibinfo {author} {\bibfnamefont {T.}~\bibnamefont {Giamarchi}},\ }\href@noop {} {\emph {\bibinfo {title} {Quantum Physics in One Dimension}}}\ (\bibinfo  {publisher} {Clarendon Press},\ \bibinfo {address} {Oxford},\ \bibinfo {year} {2003})\BibitemShut {NoStop}%
	\bibitem [{\citenamefont {Grover}\ \emph {et~al.}(2022)\citenamefont {Grover}, \citenamefont {Bocarsly}, \citenamefont {Uri}, \citenamefont {Stepanov}, \citenamefont {Di~Battista}, \citenamefont {Roy}, \citenamefont {Xiao}, \citenamefont {Meltzer}, \citenamefont {Myasoedov}, \citenamefont {Pareek}, \citenamefont {Watanabe}, \citenamefont {Taniguchi}, \citenamefont {Yan}, \citenamefont {Stern}, \citenamefont {Berg}, \citenamefont {Efetov},\ and\ \citenamefont {Zeldov}}]{Grover2022Mosaic}%
	\BibitemOpen
	\bibfield  {author} {\bibinfo {author} {\bibfnamefont {S.}~\bibnamefont {Grover}}, \bibinfo {author} {\bibfnamefont {M.}~\bibnamefont {Bocarsly}}, \bibinfo {author} {\bibfnamefont {A.}~\bibnamefont {Uri}}, \bibinfo {author} {\bibfnamefont {P.}~\bibnamefont {Stepanov}}, \bibinfo {author} {\bibfnamefont {G.}~\bibnamefont {Di~Battista}}, \bibinfo {author} {\bibfnamefont {I.}~\bibnamefont {Roy}}, \bibinfo {author} {\bibfnamefont {J.}~\bibnamefont {Xiao}}, \bibinfo {author} {\bibfnamefont {A.~Y.}\ \bibnamefont {Meltzer}}, \bibinfo {author} {\bibfnamefont {Y.}~\bibnamefont {Myasoedov}}, \bibinfo {author} {\bibfnamefont {K.}~\bibnamefont {Pareek}}, \bibinfo {author} {\bibfnamefont {K.}~\bibnamefont {Watanabe}}, \bibinfo {author} {\bibfnamefont {T.}~\bibnamefont {Taniguchi}}, \bibinfo {author} {\bibfnamefont {B.}~\bibnamefont {Yan}}, \bibinfo {author} {\bibfnamefont {A.}~\bibnamefont {Stern}}, \bibinfo {author} {\bibfnamefont {E.}~\bibnamefont {Berg}}, \bibinfo {author} {\bibfnamefont {D.~K.}\ \bibnamefont
			{Efetov}},\ and\ \bibinfo {author} {\bibfnamefont {E.}~\bibnamefont {Zeldov}},\ }\bibfield  {title} {\bibinfo {title} {{C}hern mosaic and {B}erry-curvature magnetism in magic-angle graphene},\ }\href {https://doi.org/10.1038/s41567-022-01635-7} {\bibfield  {journal} {\bibinfo  {journal} {Nat. Phys.}\ }\textbf {\bibinfo {volume} {18}},\ \bibinfo {pages} {885} (\bibinfo {year} {2022})}\BibitemShut {NoStop}%
	\bibitem [{\citenamefont {Ji}\ \emph {et~al.}(2024)\citenamefont {Ji}, \citenamefont {Park}, \citenamefont {Barber}, \citenamefont {Hu}, \citenamefont {Watanabe}, \citenamefont {Taniguchi}, \citenamefont {Chu}, \citenamefont {Xu},\ and\ \citenamefont {Shen}}]{ji2024localprobebulkedge}%
	\BibitemOpen
	\bibfield  {author} {\bibinfo {author} {\bibfnamefont {Z.}~\bibnamefont {Ji}}, \bibinfo {author} {\bibfnamefont {H.}~\bibnamefont {Park}}, \bibinfo {author} {\bibfnamefont {M.~E.}\ \bibnamefont {Barber}}, \bibinfo {author} {\bibfnamefont {C.}~\bibnamefont {Hu}}, \bibinfo {author} {\bibfnamefont {K.}~\bibnamefont {Watanabe}}, \bibinfo {author} {\bibfnamefont {T.}~\bibnamefont {Taniguchi}}, \bibinfo {author} {\bibfnamefont {J.-H.}\ \bibnamefont {Chu}}, \bibinfo {author} {\bibfnamefont {X.}~\bibnamefont {Xu}},\ and\ \bibinfo {author} {\bibfnamefont {Z.-X.}\ \bibnamefont {Shen}},\ }\bibfield  {title} {\bibinfo {title} {Local probe of bulk and edge states in a fractional {C}hern insulator},\ }\href {https://doi.org/10.1038/s41586-024-08092-7} {\bibfield  {journal} {\bibinfo  {journal} {Nature}\ }\textbf {\bibinfo {volume} {635}},\ \bibinfo {pages} {578} (\bibinfo {year} {2024})}\BibitemShut {NoStop}%
	\bibitem [{\citenamefont {Han}\ \emph {et~al.}(2024)\citenamefont {Han}, \citenamefont {Lu}, \citenamefont {Yao}, \citenamefont {Shi}, \citenamefont {Yang}, \citenamefont {Seo}, \citenamefont {Ye}, \citenamefont {Wu}, \citenamefont {Zhou}, \citenamefont {Liu}, \citenamefont {Shi}, \citenamefont {Hua}, \citenamefont {Watanabe}, \citenamefont {Taniguchi}, \citenamefont {Xiong}, \citenamefont {Fu},\ and\ \citenamefont {Ju}}]{Han_Lu_Yao_Shi2024}%
	\BibitemOpen
	\bibfield  {author} {\bibinfo {author} {\bibfnamefont {T.}~\bibnamefont {Han}}, \bibinfo {author} {\bibfnamefont {Z.}~\bibnamefont {Lu}}, \bibinfo {author} {\bibfnamefont {Y.}~\bibnamefont {Yao}}, \bibinfo {author} {\bibfnamefont {L.}~\bibnamefont {Shi}}, \bibinfo {author} {\bibfnamefont {J.}~\bibnamefont {Yang}}, \bibinfo {author} {\bibfnamefont {J.}~\bibnamefont {Seo}}, \bibinfo {author} {\bibfnamefont {S.}~\bibnamefont {Ye}}, \bibinfo {author} {\bibfnamefont {Z.}~\bibnamefont {Wu}}, \bibinfo {author} {\bibfnamefont {M.}~\bibnamefont {Zhou}}, \bibinfo {author} {\bibfnamefont {H.}~\bibnamefont {Liu}}, \bibinfo {author} {\bibfnamefont {G.}~\bibnamefont {Shi}}, \bibinfo {author} {\bibfnamefont {Z.}~\bibnamefont {Hua}}, \bibinfo {author} {\bibfnamefont {K.}~\bibnamefont {Watanabe}}, \bibinfo {author} {\bibfnamefont {T.}~\bibnamefont {Taniguchi}}, \bibinfo {author} {\bibfnamefont {P.}~\bibnamefont {Xiong}}, \bibinfo {author} {\bibfnamefont {L.}~\bibnamefont {Fu}},\ and\ \bibinfo {author} {\bibfnamefont
			{L.}~\bibnamefont {Ju}},\ }\bibfield  {title} {\bibinfo {title} {Signatures of chiral superconductivity in rhombohedral graphene},\ }\Eprint {https://arxiv.org/abs/2408.15233} {arXiv:2408.15233 [cond-mat.mes-hall]}  (\bibinfo {year} {2024})\BibitemShut {NoStop}%
	\bibitem [{\citenamefont {Jain}(1989)}]{PhysRevLett.63.199}%
	\BibitemOpen
	\bibfield  {author} {\bibinfo {author} {\bibfnamefont {J.~K.}\ \bibnamefont {Jain}},\ }\bibfield  {title} {\bibinfo {title} {Composite-fermion approach for the fractional quantum {H}all effect},\ }\href {https://doi.org/10.1103/PhysRevLett.63.199} {\bibfield  {journal} {\bibinfo  {journal} {Phys. Rev. Lett.}\ }\textbf {\bibinfo {volume} {63}},\ \bibinfo {pages} {199} (\bibinfo {year} {1989})}\BibitemShut {NoStop}%
	\bibitem [{\citenamefont {Jain}(2020)}]{jainthirty}%
	\BibitemOpen
	\bibfield  {author} {\bibinfo {author} {\bibfnamefont {J.~K.}\ \bibnamefont {Jain}},\ }\bibinfo {title} {Thirty years of composite fermions and beyond},\ in\ \href {https://doi.org/10.1142/9789811217494_0001} {\emph {\bibinfo {booktitle} {Fractional Quantum Hall Effects: New Developments}}},\ \bibinfo {editor} {edited by\ \bibinfo {editor} {\bibfnamefont {B.~I.}\ \bibnamefont {Halperin}}\ and\ \bibinfo {editor} {\bibfnamefont {J.~K.}\ \bibnamefont {Jain}}}\ (\bibinfo  {publisher} {World Scientific},\ \bibinfo {address} {New York},\ \bibinfo {year} {2020})\ Chap.~\bibinfo {chapter} {1}, pp.\ \bibinfo {pages} {1--78}\BibitemShut {NoStop}%
	\bibitem [{\citenamefont {Villegas~Rosales}\ \emph {et~al.}(2021)\citenamefont {Villegas~Rosales}, \citenamefont {Madathil}, \citenamefont {Chung}, \citenamefont {Pfeiffer}, \citenamefont {West}, \citenamefont {Baldwin},\ and\ \citenamefont {Shayegan}}]{PhysRevLett.127.056801}%
	\BibitemOpen
	\bibfield  {author} {\bibinfo {author} {\bibfnamefont {K.~A.}\ \bibnamefont {Villegas~Rosales}}, \bibinfo {author} {\bibfnamefont {P.~T.}\ \bibnamefont {Madathil}}, \bibinfo {author} {\bibfnamefont {Y.~J.}\ \bibnamefont {Chung}}, \bibinfo {author} {\bibfnamefont {L.~N.}\ \bibnamefont {Pfeiffer}}, \bibinfo {author} {\bibfnamefont {K.~W.}\ \bibnamefont {West}}, \bibinfo {author} {\bibfnamefont {K.~W.}\ \bibnamefont {Baldwin}},\ and\ \bibinfo {author} {\bibfnamefont {M.}~\bibnamefont {Shayegan}},\ }\bibfield  {title} {\bibinfo {title} {Fractional quantum {H}all effect energy gaps: Role of electron layer thickness},\ }\href {https://doi.org/10.1103/PhysRevLett.127.056801} {\bibfield  {journal} {\bibinfo  {journal} {Phys. Rev. Lett.}\ }\textbf {\bibinfo {volume} {127}},\ \bibinfo {pages} {056801} (\bibinfo {year} {2021})}\BibitemShut {NoStop}%
	\bibitem [{\citenamefont {Villegas~Rosales}\ \emph {et~al.}(2022)\citenamefont {Villegas~Rosales}, \citenamefont {Madathil}, \citenamefont {Chung}, \citenamefont {Pfeiffer}, \citenamefont {West}, \citenamefont {Baldwin},\ and\ \citenamefont {Shayegan}}]{PhysRevB.106.L041301}%
	\BibitemOpen
	\bibfield  {author} {\bibinfo {author} {\bibfnamefont {K.~A.}\ \bibnamefont {Villegas~Rosales}}, \bibinfo {author} {\bibfnamefont {P.~T.}\ \bibnamefont {Madathil}}, \bibinfo {author} {\bibfnamefont {Y.~J.}\ \bibnamefont {Chung}}, \bibinfo {author} {\bibfnamefont {L.~N.}\ \bibnamefont {Pfeiffer}}, \bibinfo {author} {\bibfnamefont {K.~W.}\ \bibnamefont {West}}, \bibinfo {author} {\bibfnamefont {K.~W.}\ \bibnamefont {Baldwin}},\ and\ \bibinfo {author} {\bibfnamefont {M.}~\bibnamefont {Shayegan}},\ }\bibfield  {title} {\bibinfo {title} {Composite fermion mass: Experimental measurements in ultrahigh quality two-dimensional electron systems},\ }\href {https://doi.org/10.1103/PhysRevB.106.L041301} {\bibfield  {journal} {\bibinfo  {journal} {Phys. Rev. B}\ }\textbf {\bibinfo {volume} {106}},\ \bibinfo {pages} {L041301} (\bibinfo {year} {2022})}\BibitemShut {NoStop}%
	\bibitem [{\citenamefont {Wu}\ \emph {et~al.}(2012)\citenamefont {Wu}, \citenamefont {Jain},\ and\ \citenamefont {Sun}}]{PhysRevB.86.165129}%
	\BibitemOpen
	\bibfield  {author} {\bibinfo {author} {\bibfnamefont {Y.-H.}\ \bibnamefont {Wu}}, \bibinfo {author} {\bibfnamefont {J.~K.}\ \bibnamefont {Jain}},\ and\ \bibinfo {author} {\bibfnamefont {K.}~\bibnamefont {Sun}},\ }\bibfield  {title} {\bibinfo {title} {Adiabatic continuity between {H}ofstadter and {C}hern insulator states},\ }\href {https://doi.org/10.1103/PhysRevB.86.165129} {\bibfield  {journal} {\bibinfo  {journal} {Phys. Rev. B}\ }\textbf {\bibinfo {volume} {86}},\ \bibinfo {pages} {165129} (\bibinfo {year} {2012})}\BibitemShut {NoStop}%
	\bibitem [{\citenamefont {Lu}\ \emph {et~al.}(2024{\natexlab{b}})\citenamefont {Lu}, \citenamefont {Chen}, \citenamefont {Wu}, \citenamefont {Sun},\ and\ \citenamefont {Meng}}]{PhysRevLett.132.236502}%
	\BibitemOpen
	\bibfield  {author} {\bibinfo {author} {\bibfnamefont {H.}~\bibnamefont {Lu}}, \bibinfo {author} {\bibfnamefont {B.-B.}\ \bibnamefont {Chen}}, \bibinfo {author} {\bibfnamefont {H.-Q.}\ \bibnamefont {Wu}}, \bibinfo {author} {\bibfnamefont {K.}~\bibnamefont {Sun}},\ and\ \bibinfo {author} {\bibfnamefont {Z.~Y.}\ \bibnamefont {Meng}},\ }\bibfield  {title} {\bibinfo {title} {Thermodynamic response and neutral excitations in integer and fractional quantum anomalous {H}all states emerging from correlated flat bands},\ }\href {https://doi.org/10.1103/PhysRevLett.132.236502} {\bibfield  {journal} {\bibinfo  {journal} {Phys. Rev. Lett.}\ }\textbf {\bibinfo {volume} {132}},\ \bibinfo {pages} {236502} (\bibinfo {year} {2024}{\natexlab{b}})}\BibitemShut {NoStop}%
	\bibitem [{\citenamefont {Xie}\ and\ \citenamefont {Das~Sarma}(2024)}]{PhysRevB.109.L241115}%
	\BibitemOpen
	\bibfield  {author} {\bibinfo {author} {\bibfnamefont {M.}~\bibnamefont {Xie}}\ and\ \bibinfo {author} {\bibfnamefont {S.}~\bibnamefont {Das~Sarma}},\ }\bibfield  {title} {\bibinfo {title} {Integer and fractional quantum anomalous {H}all effects in pentalayer graphene},\ }\href {https://doi.org/10.1103/PhysRevB.109.L241115} {\bibfield  {journal} {\bibinfo  {journal} {Phys. Rev. B}\ }\textbf {\bibinfo {volume} {109}},\ \bibinfo {pages} {L241115} (\bibinfo {year} {2024})}\BibitemShut {NoStop}%
	\bibitem [{\citenamefont {Hu}\ \emph {et~al.}(2009)\citenamefont {Hu}, \citenamefont {Rezayi}, \citenamefont {Wan},\ and\ \citenamefont {Yang}}]{PhysRevB.80.235330}%
	\BibitemOpen
	\bibfield  {author} {\bibinfo {author} {\bibfnamefont {Z.-X.}\ \bibnamefont {Hu}}, \bibinfo {author} {\bibfnamefont {E.~H.}\ \bibnamefont {Rezayi}}, \bibinfo {author} {\bibfnamefont {X.}~\bibnamefont {Wan}},\ and\ \bibinfo {author} {\bibfnamefont {K.}~\bibnamefont {Yang}},\ }\bibfield  {title} {\bibinfo {title} {Edge-mode velocities and thermal coherence of quantum {H}all interferometers},\ }\href {https://doi.org/10.1103/PhysRevB.80.235330} {\bibfield  {journal} {\bibinfo  {journal} {Phys. Rev. B}\ }\textbf {\bibinfo {volume} {80}},\ \bibinfo {pages} {235330} (\bibinfo {year} {2009})}\BibitemShut {NoStop}%
	\bibitem [{\citenamefont {Thomson}\ and\ \citenamefont {Alicea}(2021)}]{thomson2021:PhysRevB.103.125138}%
	\BibitemOpen
	\bibfield  {author} {\bibinfo {author} {\bibfnamefont {A.}~\bibnamefont {Thomson}}\ and\ \bibinfo {author} {\bibfnamefont {J.}~\bibnamefont {Alicea}},\ }\bibfield  {title} {\bibinfo {title} {Recovery of massless {D}irac fermions at charge neutrality in strongly interacting twisted bilayer graphene with disorder},\ }\href {https://doi.org/10.1103/PhysRevB.103.125138} {\bibfield  {journal} {\bibinfo  {journal} {Phys. Rev. B}\ }\textbf {\bibinfo {volume} {103}},\ \bibinfo {pages} {125138} (\bibinfo {year} {2021})}\BibitemShut {NoStop}%
	\bibitem [{\citenamefont {Shavit}\ and\ \citenamefont {Oreg}(2022)}]{shavit2022:PhysRevLett.128.156801}%
	\BibitemOpen
	\bibfield  {author} {\bibinfo {author} {\bibfnamefont {G.}~\bibnamefont {Shavit}}\ and\ \bibinfo {author} {\bibfnamefont {Y.}~\bibnamefont {Oreg}},\ }\bibfield  {title} {\bibinfo {title} {Domain formation driven by the entropy of topological edge modes},\ }\href {https://doi.org/10.1103/PhysRevLett.128.156801} {\bibfield  {journal} {\bibinfo  {journal} {Phys. Rev. Lett.}\ }\textbf {\bibinfo {volume} {128}},\ \bibinfo {pages} {156801} (\bibinfo {year} {2022})}\BibitemShut {NoStop}%
	\bibitem [{\citenamefont {Wen}(2004)}]{WenBook}%
	\BibitemOpen
	\bibfield  {author} {\bibinfo {author} {\bibfnamefont {X.-G.}\ \bibnamefont {Wen}},\ }\href@noop {} {\emph {\bibinfo {title} {Quantum Field Theory of Many-Body Systems: From the Origin of Sound to an Origin of Light and Electrons}}}\ (\bibinfo  {publisher} {Oxford University Press},\ \bibinfo {address} {Oxford},\ \bibinfo {year} {2004})\BibitemShut {NoStop}%
	\bibitem [{\citenamefont {Ferraro}\ and\ \citenamefont {Sukhorukov}(2017)}]{ferraro2017:10.21468/SciPostPhys.3.2.014}%
	\BibitemOpen
	\bibfield  {author} {\bibinfo {author} {\bibfnamefont {D.}~\bibnamefont {Ferraro}}\ and\ \bibinfo {author} {\bibfnamefont {E.}~\bibnamefont {Sukhorukov}},\ }\bibfield  {title} {\bibinfo {title} {Interaction effects in a multi-channel {F}abry-{P\'e}rot interferometer in the {A}haronov-{B}ohm regime},\ }\href {https://doi.org/10.21468/SciPostPhys.3.2.014} {\bibfield  {journal} {\bibinfo  {journal} {SciPost Phys.}\ }\textbf {\bibinfo {volume} {3}},\ \bibinfo {pages} {014} (\bibinfo {year} {2017})}\BibitemShut {NoStop}%
	\bibitem [{\citenamefont {Wei}\ \emph {et~al.}(2024)\citenamefont {Wei}, \citenamefont {Feldman},\ and\ \citenamefont {Halperin}}]{wei2024:PhysRevB.110.075306}%
	\BibitemOpen
	\bibfield  {author} {\bibinfo {author} {\bibfnamefont {Z.}~\bibnamefont {Wei}}, \bibinfo {author} {\bibfnamefont {D.~E.}\ \bibnamefont {Feldman}},\ and\ \bibinfo {author} {\bibfnamefont {B.~I.}\ \bibnamefont {Halperin}},\ }\bibfield  {title} {\bibinfo {title} {Quantum {H}all interferometry at finite bias with multiple edge channels},\ }\href {https://doi.org/10.1103/PhysRevB.110.075306} {\bibfield  {journal} {\bibinfo  {journal} {Phys. Rev. B}\ }\textbf {\bibinfo {volume} {110}},\ \bibinfo {pages} {075306} (\bibinfo {year} {2024})}\BibitemShut {NoStop}%
	\bibitem [{\citenamefont {Imry}\ and\ \citenamefont {Ma}(1975)}]{imry1975:PhysRevLett.35.1399}%
	\BibitemOpen
	\bibfield  {author} {\bibinfo {author} {\bibfnamefont {Y.}~\bibnamefont {Imry}}\ and\ \bibinfo {author} {\bibfnamefont {S.-k.}\ \bibnamefont {Ma}},\ }\bibfield  {title} {\bibinfo {title} {Random-field instability of the ordered state of continuous symmetry},\ }\href {https://doi.org/10.1103/PhysRevLett.35.1399} {\bibfield  {journal} {\bibinfo  {journal} {Phys. Rev. Lett.}\ }\textbf {\bibinfo {volume} {35}},\ \bibinfo {pages} {1399} (\bibinfo {year} {1975})}\BibitemShut {NoStop}%
	\bibitem [{\citenamefont {Shavit}(2024)}]{Shavit_2024}%
	\BibitemOpen
	\bibfield  {author} {\bibinfo {author} {\bibfnamefont {G.}~\bibnamefont {Shavit}},\ }\bibfield  {title} {\bibinfo {title} {Entropy-enhanced fractional quantum anomalous {H}all effect},\ }\href {https://doi.org/10.1103/PhysRevB.110.L201406} {\bibfield  {journal} {\bibinfo  {journal} {Phys. Rev. B}\ }\textbf {\bibinfo {volume} {110}},\ \bibinfo {pages} {L201406} (\bibinfo {year} {2024})}\BibitemShut {NoStop}%
	\bibitem [{\citenamefont {Wan}\ \emph {et~al.}(2023)\citenamefont {Wan}, \citenamefont {Sarkar}, \citenamefont {Lin},\ and\ \citenamefont {Sun}}]{PhysRevLett.130.216401}%
	\BibitemOpen
	\bibfield  {author} {\bibinfo {author} {\bibfnamefont {X.}~\bibnamefont {Wan}}, \bibinfo {author} {\bibfnamefont {S.}~\bibnamefont {Sarkar}}, \bibinfo {author} {\bibfnamefont {S.-Z.}\ \bibnamefont {Lin}},\ and\ \bibinfo {author} {\bibfnamefont {K.}~\bibnamefont {Sun}},\ }\bibfield  {title} {\bibinfo {title} {Topological exact flat bands in two-dimensional materials under periodic strain},\ }\href {https://doi.org/10.1103/PhysRevLett.130.216401} {\bibfield  {journal} {\bibinfo  {journal} {Phys. Rev. Lett.}\ }\textbf {\bibinfo {volume} {130}},\ \bibinfo {pages} {216401} (\bibinfo {year} {2023})}\BibitemShut {NoStop}%
	\bibitem [{\citenamefont {Wu}\ \emph {et~al.}(2024)\citenamefont {Wu}, \citenamefont {Sarkar}, \citenamefont {Wan}, \citenamefont {Sun},\ and\ \citenamefont {Lin}}]{wu2024quantum}%
	\BibitemOpen
	\bibfield  {author} {\bibinfo {author} {\bibfnamefont {A.-K.}\ \bibnamefont {Wu}}, \bibinfo {author} {\bibfnamefont {S.}~\bibnamefont {Sarkar}}, \bibinfo {author} {\bibfnamefont {X.}~\bibnamefont {Wan}}, \bibinfo {author} {\bibfnamefont {K.}~\bibnamefont {Sun}},\ and\ \bibinfo {author} {\bibfnamefont {S.-Z.}\ \bibnamefont {Lin}},\ }\bibfield  {title} {\bibinfo {title} {Quantum-metric-induced quantum {H}all conductance inversion and reentrant transition in fractional {C}hern insulators},\ }\href {https://doi.org/10.1103/PhysRevResearch.6.L032063} {\bibfield  {journal} {\bibinfo  {journal} {Phys. Rev. Res.}\ }\textbf {\bibinfo {volume} {6}},\ \bibinfo {pages} {L032063} (\bibinfo {year} {2024})}\BibitemShut {NoStop}%
	\bibitem [{\citenamefont {Haldane}(1988)}]{PhysRevLett.61.2015}%
	\BibitemOpen
	\bibfield  {author} {\bibinfo {author} {\bibfnamefont {F.~D.~M.}\ \bibnamefont {Haldane}},\ }\bibfield  {title} {\bibinfo {title} {Model for a quantum {H}all effect without {L}andau levels: Condensed-matter realization of the ``parity anomaly''},\ }\href {https://doi.org/10.1103/PhysRevLett.61.2015} {\bibfield  {journal} {\bibinfo  {journal} {Phys. Rev. Lett.}\ }\textbf {\bibinfo {volume} {61}},\ \bibinfo {pages} {2015} (\bibinfo {year} {1988})}\BibitemShut {NoStop}%
\end{thebibliography}
\end{document}